\documentclass[aoas,preprint]{imsart}

\usepackage{amsmath, amssymb}
\allowdisplaybreaks[1]
\usepackage{amsthm}
\usepackage{amsfonts}

\usepackage{graphicx}
\usepackage{psfrag,epsf}
\usepackage{enumerate}
\usepackage{natbib}
\usepackage{url} 
\usepackage{bm}
\RequirePackage[colorlinks,citecolor=blue,urlcolor=blue]{hyperref}
\usepackage{floatrow}
\DeclareFloatFont{small}{\small}
\usepackage[font={footnotesize}]{caption}

\RequirePackage[OT1]{fontenc}
\RequirePackage{amsthm,amsmath}
\RequirePackage{natbib}
\RequirePackage[colorlinks,citecolor=blue,urlcolor=blue]{hyperref}

\newcommand{\Prob}{{\rm I}\kern-0.18em{\rm P}}
\newcommand{\1}{{\rm 1}\kern-0.24em{\rm I}}
\DeclareMathOperator*{\argmin}{arg\!min}
\DeclareMathOperator*{\argmax}{arg\!max}

\usepackage[usenames, dvipsnames]{color}
\newcommand{\WL}[1]{\textcolor{black}{#1}}
\newcommand{\JL}[1]{\textcolor{black}{#1}}

\newcommand{\ww}[1]{\textcolor{black}{#1}}
\newcommand{\Rtwo}[1]{\textcolor{black}{#1}}
\arxiv{arXiv:1603.05915}

\startlocaldefs
\numberwithin{equation}{section}
\theoremstyle{plain}

\newtheorem{lemma}{Lemma}[section]
\endlocaldefs

\usepackage{pdfpages}

\begin{document}

\begin{frontmatter}
\title{MSIQ: Joint Modeling of Multiple RNA-seq Samples for Accurate Isoform Quantification}
\runtitle{Isoform Quantification on Multiple RNA-seq Samples}

\begin{aug}
\author{\fnms{Wei Vivian} \snm{Li}\thanksref{m1,t1}\ead[label=e1]{liw@ucla.edu}},
\author{\fnms{Anqi} \snm{Zhao}\thanksref{m2}\ead[label=e2]{anqizhao@fas.harvard.edu}},
\author{\fnms{Shihua} \snm{Zhang}\thanksref{m3,t2}\ead[label=e3]{zsh@amss.ac.cn}}\\
\and
\author{\fnms{Jingyi Jessica} \snm{Li}\thanksref{m1,t1,t2}\ead[label=e4]{jli@stat.ucla.edu}}

\thankstext{t1}{Equal contribution.}
\thankstext{t2}{Corresponding authors. Please send email correspondence to jli@stat.ucla.edu or zsh@amss.ac.cn.}
\runauthor{W. Li et al.}

\affiliation{University of California, Los Angeles\thanksmark{m1}, Harvard University\thanksmark{m2} and Chinese Academy of Sciences\thanksmark{m3}}

\address{Wei Vivian Li\\
Department of Statistics\\
8125 Math Sciences Bldg.\\
University of California, Los Angeles\\
Los Angeles, CA 90095-1554\\
\printead{e1}\\
\phantom{E-mail:\ }}

\address{Anqi Zhao\\
Department of Statistics\\
Science Center 7th floor\\
One Oxford Street\\
Harvard University\\    
Cambridge, MA 02138-2901\\
\printead{e2}\\
\phantom{E-mail:\ }}

\address{Shihua Zhang\\
Institute of Applied Mathematics\\ 
Academy of Mathematics and Systems Science\\
Chinese Academy of Sciences\\
No.55, Zhongguancun East road\\
Beijing 100190, China\\
\printead{e3}\\
\\
\phantom{E-mail:\ }}

\address{Jingyi Jessica Li\\
Department of Statistics\\
8125 Math Sciences Bldg.\\
University of California, Los Angeles\\
Los Angeles, CA 90095-1554\\
\printead{e4}\\
\phantom{E-mail:\ }}
\end{aug}

\begin{abstract}
Next-generation RNA sequencing (RNA-seq) technology has been widely used to assess full-length RNA isoform abundance in a high-throughput manner. RNA-seq data offer insight into gene expression levels and transcriptome structures, enabling us to better understand the regulation of gene expression and fundamental biological processes.
Accurate isoform quantification from RNA-seq data is \ww{challenging} due to the information loss in sequencing experiments. \ww{A recent} accumulation of multiple RNA-seq data sets 
\ww{from the same tissue or cell type}
provides new opportunities to improve the \ww{accuracy of isoform quantification}.
However, existing statistical or computational methods for multiple RNA-seq samples either pool the samples into one sample or assign equal weights to the samples \ww{when} estimating isoform abundance.
These methods ignore the possible heterogeneity in the quality 
of different samples \ww{and could result in} biased and unrobust estimates.
In this article, we develop a method\ww{, which we call} ``joint modeling of multiple RNA-seq samples for accurate isoform quantification'' (MSIQ)\ww{,} for more accurate and robust isoform quantification \ww{by} integrating multiple RNA-seq samples under a Bayesian framework.
Our method aims to (1) identify \ww{a} consistent group of samples with homogeneous quality and (2) improve isoform quantification accuracy by jointly modeling multiple RNA-seq samples \ww{by allowing for higher} weights on the consistent group.
We show that MSIQ provides a consistent estimator of isoform abundance, and \ww{we} demonstrate the accuracy and effectiveness of MSIQ compared \ww{with} alternative methods through simulation studies on \emph{D. melanogaster} genes. We justify MSIQ's advantages over existing approaches via application studies on real RNA-seq data \ww{from} human embryonic stem cells, brain tissues, and \ww{the} HepG2 immortalized cell line. We also perform a comprehensive analysis \ww{of} how the isoform quantification accuracy would be affected by RNA-seq sample heterogeneity and different experimental protocols.
\end{abstract}



\end{frontmatter}

\section{Introduction}
\label{sec:intro}
Transcriptomes are complete sets of RNA molecules in biological samples. Unlike the genome\ww{, which} is largely invariant in different tissues and cells of the same individual, transcriptomes can vary greatly and cause different tissue and cell phenotypes. Understanding transcriptomes is essential for interpreting genome \ww{function} and investigating molecular bases for various disease phenomena.
In transcriptomes, the most important components are messenger RNA (mRNA) transcripts, as they will be translated into proteins---the key functional units in most biological processes.
During the transcription process from genes to mRNA transcripts, one gene may give rise to multiple mRNA transcripts with different nucleotide sequences, thus \ww{contributing to} the diversity of transcriptomes. 
mRNA transcripts from the same gene are often referred to as \textit{isoforms}, which are different combinations of whole or partial \textit{exons} (i.e., \ww{contiguous genomic regions within} genes that will be transcribed into RNA molecules).

Transcriptomics is an emerging field and one of its primary goals is to quantify the dynamic expression levels of mRNA isoforms under different biological conditions.
For common species (e.g., \ww{\textit{Homo sapiens} (humans), \textit{Mus musculus} (mice), \textit{Drosophila melanogaster} (fruit flies)}, etc.), extant gene annotations record a large number of mRNA isoforms reported \ww{in} previous literature. 
For example, \ww{the} UCSC genome browser \citep{kent2002human}, GENCODE \citep{harrow2012gencode} and RefSeq \citep{pruitt2014refseq} contain known mRNA isoform structures in transcriptomes of \ww{humans} and several other species.
However, the annotations lack gold standard abundance information of these isoforms. In many biological studies, \ww{it is important to identify and catalog expression levels of mRNA isoforms \citep{hansen2011sequencing} in order to perform downstream analyses such as identification of differentially expressed genes and construction of transcript co-expression networks.}
Hence, how to accurately estimate isoform abundance \ww{is} a key question.

Over the past decade, next-generation RNA sequencing (RNA-seq) technologies have generated numerous data sets with unprecedented nucleotide-level information  on transcriptomes, providing new opportunities to study \ww{the} dynamic expression of known and novel mRNAs in a high-throughput manner \citep{wang2009rna,conesa2016survey,trapnell2009tophat}.
\ww{The ideal data would include the sequences of full-length mRNA transcripts; however,} most widely used next-generation Illumina sequencers generate millions of short sequences called \textit{reads} (typically shorter than $400$ base pairs) from the two ends of mRNA transcript fragments \citep{wang2009rna}, \ww{while} other third-generation sequencing technologies (e.g., Ion Torrent and Pacific Biosciences) produce longer but more erroneous reads \citep{quail2012tale}. In this paper, our discussion focuses on \WL{paired-end} RNA-seq data generated by Illumina sequencers. 
For more details \ww{on}  Illumina RNA-seq \ww{experiments}, see \ww{Supplementary} Fig S1.

\begin{figure}[tbh!]
\begin{center}
\includegraphics[width=0.8\textwidth]{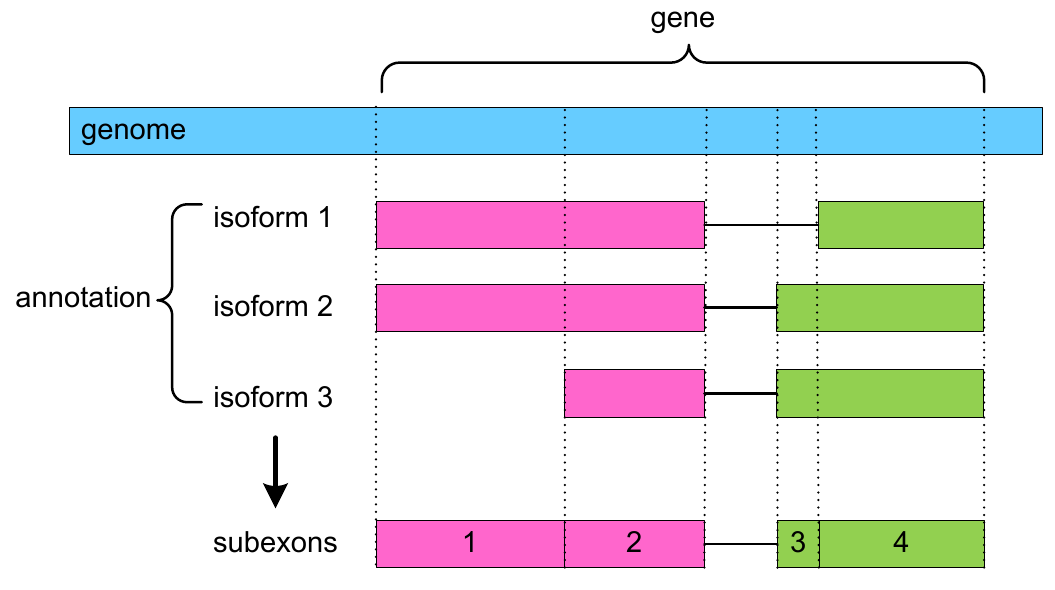}
\end{center}
\caption{Definition of subexons. The example gene has two exons, represented by magenta and green boxes, and three mRNA isoforms. The solid lines between exons represent introns in the gene that have been spliced out in isoforms. Adjacent splicing sites in these isoforms define four non-overlapping subexons: the first exon is divided into subexon 1 and 2, and the second exon is divided into subexon 3 and 4. 
\label{fig:subexon}}
\end{figure}

\ww{Due to the presence of} numerous isoforms in existing annotations, \ww{inference on} their abundance from RNA-seq reads has been an active \ww{field of research} since 2009 \citep{jiang2009statistical, trapnell2010transcript, li2011sparse, zhang2014wemiq}. A necessary  step is to first map (or align) reads to reference genomes so that researchers know the numbers of reads generated from each exon. \ww{Then,} a common approach to summarize RNA-seq reads is to categorize the reads by the genomic regions \ww{to which they map} so that the number of reads in different genomic regions can be used to distinguish the abundance of \ww{various} isoforms.
As different isoforms may consist of overlapping but not identical exons, many methods divide exons into \textit{subexons}, which are defined as transcribed regions between every two adjacent splicing sites in annotations \citep{li2011sparse, zhang2014wemiq, ye2016nmfp}. By this definition, every gene is composed of non-overlapping subexons and introns (i.e., non-transcribed genomic regions). 
 In Fig \ref{fig:subexon}, we illustrate a toy example of a gene with three annotated isoforms and four subexons.
Because combinations of subexons form a superset of all the annotated isoforms, it is reasonable to categorize RNA-seq reads based on the sets of subexons \ww{to which they map}. For the ease of terminology, we will refer to subexons as exons \ww{for} the remainder of this paper.
For more details \ww{regarding} categorizing RNA-seq reads, see Section \ref{sec:bin}.

\begin{figure}[tbh!]
\begin{center}
\includegraphics[trim= 2cm 21cm 4cm 1.7cm, width=0.7\textwidth]{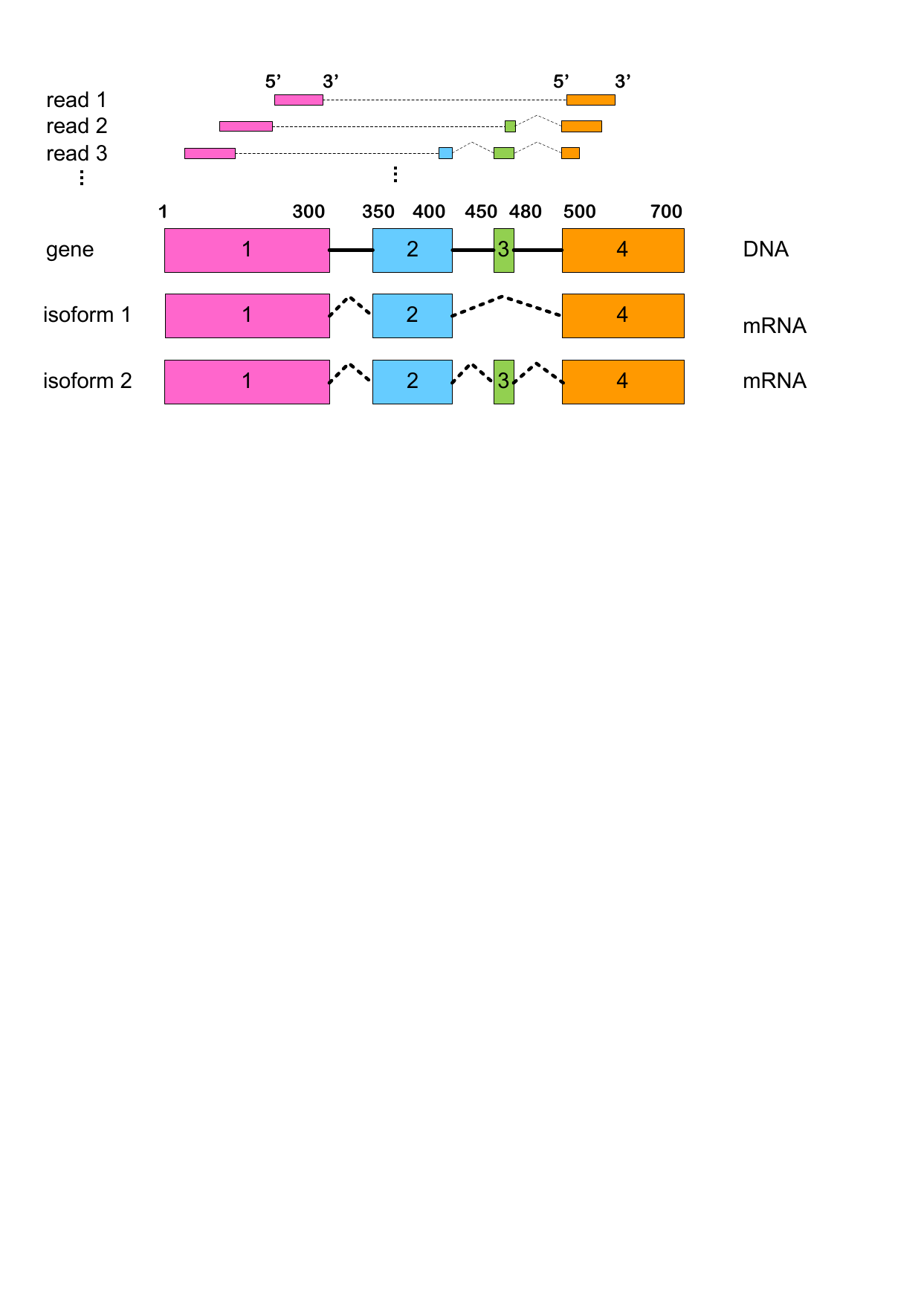}
\end{center}
\caption{Illustration of RNA-seq read generation from a hypothetical gene. The four exons of this gene are represented as boxes of different lengths and colors. The starting and \ww{ending} positions of the four exons are marked on top of the gene. In an RNA-seq experiment, multiple reads are generated and the number of reads coming from each isoform is proportional to the isoform's abundance. Each read has a 5'-end and \ww{a} 3'-end, as shown in read 1. These reads are mapped to the reference genome and their overlapping exons are key information for estimating isoform abundance.
\label{fig:intro_gene}}
\end{figure}

How to infer isoform abundance from observed RNA-seq reads is a statistical problem, as reads are generated from a mixture of isoforms. We illustrate this using a toy example in Fig \ref{fig:intro_gene}.
A hypothetical gene is composed of four non-overlapping exons.
Suppose that the gene is transcribed into two mRNA isoforms: 60\% of the transcripts are isoform 1, which consists of exons 1, 2 and 4, and 40\% of the transcripts are isoform 2, which consists of all four exons. 
In reality, the isoform proportions, though of great interest to biologists, remain unobservable under the current experimental settings. Our aim is to estimate the  relative abundance of annotated isoforms based on reads generated in RNA-seq experiments. 
Suppose that $n$ paired-end reads are generated from mRNA transcripts of the gene, and they are mapped (or aligned) to the reference genome. 
Some of the mapped reads have obvious isoform origins. For example, read $3$ is compatible only with isoform 2, and thus  must have isoform 2 as its origin. On the other hand, many mapped reads can have ambiguous origins. For example, read 1 is compatible with both isoforms 1 and 2, and thus we cannot determine its origin isoform.
The much more complex structures of real genes complicate the situation even further\ww{;} human genes have \ww{nine} exons on average \citep{sakharkar2004distri}, and a large proportion of human genes have more than \ww{ten} annotated isoforms (see Supplementary Fig S2B).
Therefore, this problem requires powerful statistical methods to provide good estimates of isoform proportions.

A number of isoform quantification methods have been developed to estimate the abundance of \ww{specific} isoforms. 
\WL{These} methods perform isoform quantification using either direct computation or model-based approaches \citep{wang2009rna, steijger2013assessment, kanitz2015comparative}.
Direct computation \ww{approaches use a variety of methods} to count the number of reads compatible with each isoform and then normalize the counts by isoform lengths and \ww{the} total number of reads to \ww{generate} estimates of isoform abundance. The most commonly used unit is reads per kilobase of transcript per million mapped reads (RPKM) \citep{mortazavi2008mapping}.
However, for complex gene structures, counts of RNA-seq reads compatible with isoforms may not be proportional to isoform abundance, as multiple isoforms can share exons and some reads cannot be assigned unequivocally to only one isoform.
To address this issue, model-based approaches are needed to assess the \ww{likelihood} of a read coming from different isoforms.
In the first model-based isoform quantification method \citep{jiang2009statistical}, read counts in genomic regions are modeled as Poisson variables (with isoform abundance as the mean parameter), under the assumption that reads are uniformly sampled within each isoform. Isoform abundance is estimated by maximum likelihood estimates.
\JL{
Cufflinks \citep{trapnell2010transcript}, the most widely used method for discovering novel isoforms from RNA-seq data, also has the functionality to estimate isoform abundance. Its approach is similar to the likelihood-based approach in \cite{jiang2009statistical}, and it proposed a new unit for isoform abundance based on paired-end RNA-seq data: fragments per kilobase of transcript per million mapped reads (FPKM), which accounts for the dependency between paired-end reads. 
}
MISO \citep{katz2010analysis} is another model-based method constructed under a Bayesian framework, and it provides maximum\textit{-a-posteriori} estimates and confidence intervals of isoform abundance. 
There are other isoform quantification methods with different features \citep{pachter2011models}. For example, SLIDE \citep{li2011sparse}  uses a linear model and \ww{can be used with various} data types; 
iReckon \citep{mezlini2013ireckon} utilizes \ww{a} regularized Expectation-Maximization algorithm;
WemIQ \citep{zhang2014wemiq} \ww{replaces} the Poisson distribution with a more general and realistic generalized Poisson distribution; eXpress \citep{roberts2013streaming} is an efficient streaming method based on an online-EM algorithm and \ww{is considered to be a faster version of Cufflœinks with comparable performance}; and Sailfish \citep{patro2014sailfish} is a fast alignment-free method that saves the read mapping step.

\begin{figure}[tbh!]
\begin{center}
\includegraphics[width=\textwidth]{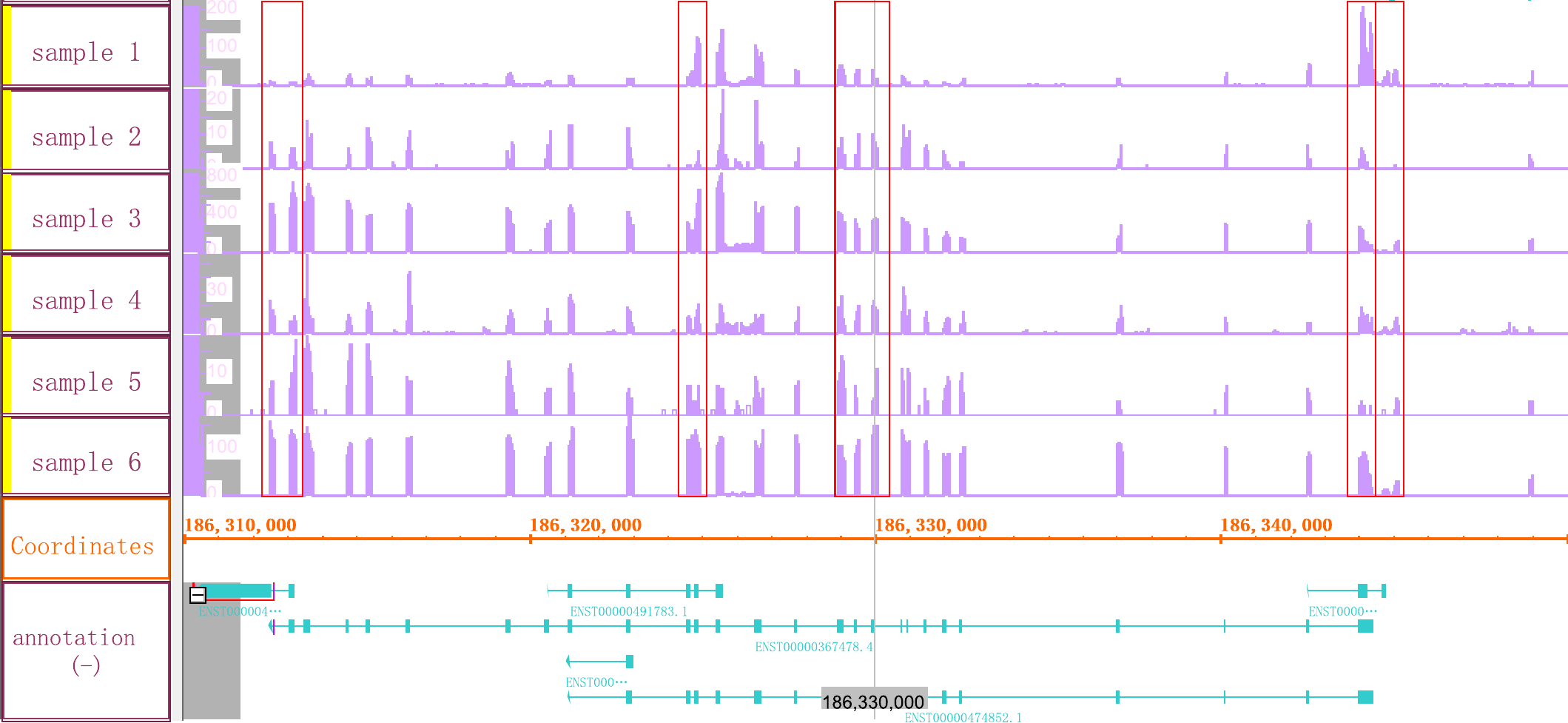}
\end{center}
\caption{
\ww{Reads in six hESC RNA-seq samples mapped to the human gene \textit{TPR}. Detailed information on these samples is listed in Supplementary Table S2. The counts of RNA-seq reads are summarized in the histograms. The annotation of the gene and isoform structures is shown in the bottom row. \Rtwo{We mark four example sites where the six samples are obviously inconsistent with red rectangles.}}
\label{fig:gene_ex}}
\end{figure}

However, there remains much space to improve the accuracy of isoform quantification due to noise and biases in RNA-seq data.
\ww{Because of the accumulation of RNA-seq samples in public databases,} multiple RNA-seq data sets are now often available for \ww{the same biological condition (e.g., the same cell or tissue type)}, and they provide more information than a single RNA-seq data set. 
\ww{For example, the GTEx (Genotype-Tissue Expression) study comprises $9,662$ samples from $54$ tissues, and the Cancer Genome Atlas (TCGA) study comprises $11,350$ samples from $33$ cancer types \citep{collado2017reproducible}.}
\ww{Here, the concept of \textit{multiple samples} includes both \textit{technical replicates}-different aliquots of the same sample measured multiple times \citep{hansen2011sequencing}- and \textit{biological replicates}-replicates obtained from multiple samples of the same material, type of cells, or tissue.}
\ww{The availability of multiple RNA-seq samples from the same biological condition (e.g., human embryonic stem cells) in public databases (e.g., NIH Gene Expression Omnibus \citep{barrett2013ncbi}) motivated us to develop a new statistical method for better isoform quantification by taking advantage of the common and thus more reliable information provided by multiple samples. The necessity of such a method is two-fold. 
First, the number of RNA-seq samples produced by a single lab is limited since experimental costs increase each time an additional replicate is added. A statistical method that allows for multiple samples enables researchers to combine their own data with public data to obtain more accurate and robust isoform abundance estimates. Second, such a method supports better reuse of public data for both new biological findings and method development.}

Several methods have been developed to use multiple RNA-seq samples \ww{from} the same biological condition for isoform quantification.
For example, CLIIQ \citep{lin2012cliiq} uses integer linear programming to jointly model RNA-seq data from multiple samples. 
MITIE \citep{behr2013mitie} assumes that the same isoforms are expressed in all samples but may have different abundances, and it then reduces the problem to solving systems of linear equations.
\WL{
FlipFlop \citep{bernard2014efficient}  uses a convex formulation and introduces the group-lasso penalty to ensure sparsity in estimation.
}
However, none of these methods \ww{considers} the quality variation of different RNA-seq samples or how such variation might affect the inference \ww{of} isoform abundance.
It is commonly recognized that RNA-seq samples generated by different protocols or different labs can \ww{vary greatly with respect to the} signal-to-noise ratios, biases, etc. 
For example, Fig \ref{fig:gene_ex} shows the RNA-seq read coverage profiles of the human gene \textit{TPR} in six human embryonic stem cell (hESC) samples. There is obvious variation in the read coverage profiles of these six samples. For example,
sample 2 has little signal in the last exon while \ww{the} other samples have obviously stronger signals in the last exon. Thus\ww{,} it is inappropriate to treat all the samples equally \ww{during} isoform quantification by assuming that they come from the same population \ww{(i.e., the same tissue or cell of interest)}.
Hence, results \ww{from} these methods \ww{may} be sensitive to the heterogeneity of samples or even\ww{,} in some cases\ww{,} be dominated by biased samples, which do not accurately reflect the transcriptome information of the \WL{given tissue type}.

In this paper, we propose a robust quantification method for isoform expression: joint modeling of \textbf{M}ultiple RNA-seq \textbf{S}amples for accurate \textbf{I}soform \textbf{Q}uantification (MSIQ). 
MSIQ is a model-based approach \ww{for} estimating isoform abundance by discerning and using multiple RNA-seq samples that share similar transcriptome information, which we define as the \textit{consistent group} in this paper. Our modeling consists of two components:
(1) estimating the probability of each sample being in the consistent group via evaluating the sample similarities, and (2) estimating isoform abundance from \ww{reweighted} samples\ww{, with greater weights given to the samples that are more likely to be consistent. These two components enable the method to distinguish between the large variation stemming from experimental factors and the reasonable biological variation.}
In Section \ref{sec:meth}\ww{,} we describe the Bayesian hierarchical model used in MSIQ to bridge unknown isoform proportions and observed read counts mapped to a gene in multiple RNA-seq samples. Our model allows for different isoform proportions of RNA-seq samples \ww{inside and outside} the consistent group\ww{;} a main parameter of interest \ww{relates to} the isoform proportions in the consistent group. This approach reduces the \ww{probability} that the estimated isoform abundance \ww{is} biased by samples \ww{of poor} quality. We conduct parameter inference by Gibbs \ww{sampling and prove} the consistency of the MSIQ estimator. 
We show that the \ww{isoform proportions estimated} by MSIQ are consistent with the unknown isoform proportions in the consistent group, while the estimates based on the assumption that all samples have equal \ww{weights} are not.
In Section \ref{sec:results}\ww{,} we apply MSIQ to both simulated and real data sets to illustrate the efficiency and robustness of MSIQ under various parameter settings and \ww{with} different parameter estimation procedures. We also compare MSIQ with the oracle estimators and other \WL{widely used} estimation methods.
In Section \ref{sec:conc}, we discuss the advantages and limitations of MSIQ and its possible extensions.

\section{Methods}
\label{sec:meth}
For a given gene, our proposed MSIQ method aims \ww{to achieve two goals with respect to} isoform expression quantification.
First, we want to identify the samples \ww{that represent} the tissue or cell type of interest.  We refer to these samples as the \textit{consistent group} and assume that the group contains at least one sample. We identify samples in the consistent group under the assumption that these samples share \ww{the} most similar read distributions among all the samples.
Second, we would also like to estimate the proportion of reads coming from each mRNA isoform in the given tissue or cell type, with larger \ww{weights} given to the samples in the consistent group.
We focus our efforts on RNA-seq data with paired-end reads, but the model can easily \ww{be} extended \ww{for} single-end reads.

\subsection{Ideal and practical parameters of interest}
Suppose we are studying a  gene with $N$ exons, $J$ annotated mRNA isoforms, and $D$ RNA-seq samples.
 Ideally, we are interested in the true proportion of each isoform
$$p_j = P(\text{an mRNA transcript is of isoform}\ j),\ j=1,2,...,J.$$
However, \ww{these hidden parameters are not} observable in RNA-seq experiments, which do not directly measure full-length mRNA transcripts. Instead of directly estimating $p_j$, we aim \ww{to estimate} the practical parameters 
$$\alpha_j = P(\text{an RNA-seq read is from isoform}\ j),\ j=1,2,...,J,$$
which we refer to as isoform proportions in our discussion.

\subsection{Observed data}\label{sec:bin}

We denote the observed data, $D$ independent samples of \ww{reads mapped} to the given gene, by
$$ \bm R^{(d)} = \{r^{(d)}_{1}, r^{(d)}_{2}, \dots, r^{(d)}_{n_d}\}\,,\quad d = 1\,, 2\,, \dots, D\,,$$
where $n_d$ and $r^{(d)}_{i}$\ww{,} respectively\ww{,} denote the total number of reads and the $i$th read ($i=1,2,\dots,n_d$) in sample $d$. 
To use the read information, \ww{an} efficient data summary is needed to preserve \ww{the} most relevant information for isoform quantification while \ww{limiting} the computational complexity \ww{to} a manageable level \citep{rossell2014quantifying}. 
We write each read as
$$r^{(d)}_{i} = \left\{\bm s^{(d)}_{1i}, \bm s^{(d)}_{2i}, 
\left\{y_{i1}^{(d)}, y_{ic^{(d)}}^{(d)}, y_{i(c^{(d)}+1)}^{(d)}, y_{i(2c^{(d)})}^{(d)}\right\}\right\},$$
where $\bm s^{(d)}_{1i} \ \text{and}\ \bm s^{(d)}_{2i}$\ww{,} respectively\ww{,} denote the index set of exons overlapping with the read's left end and right end; 
$y_{ik}^{(d)}$ denotes the $k$th genomic position of read $i$; and $c^{(d)}$ is the read length in sample $d$. \WL{Please refer to the supplementary information for a more detailed discussion on the advantages of this summarizing approach \JL{over other existing approaches}.}

\subsection{Assumptions and prior}
\ww{In addition to} the observed data, \ww{we consider} the hidden data\ww{, which} are the isoform origins of the reads:
\begin{eqnarray*}
\bm Z^{(d)} &=& (Z^{(d)}_{1}, Z^{(d)}_{2}, \dots, Z^{(d)}_{n_d})^\prime\,,
\end{eqnarray*}
where $Z^{(d)}_{i} \in \{1,2,\dots,J\}$ indicates the isoform origin of read $i$, and $Z^{(d)}_{i} =j$ if read $r^{(d)}_{i}$ actually comes from isoform $j$.

\ww{The differences between RNA-seq samples are} reflected in their isoform proportion $\bm \tau^{(d)} $, $d = 1, 2, \dots, D$. In RNA-seq sample $d$, we denote the true probability of reads from isoform $j$ as $\tau^{(d)}_j= P(Z^{(d)}_{i}=j)$ and the isoform proportion vector as
\begin{eqnarray*}
\bm \tau^{(d)} = (\tau^{(d)}_1, \tau^{(d)}_2, \dots, \tau^{(d)}_J)^\prime\,,
\end{eqnarray*}
with $\sum^J_{j=1}\tau^{(d)}_j=1$.
We define a hidden state variable $E_d$ for each sample such that 
$$E_d = \1\{\text{sample $d$ belongs to the consistent group}\}.$$ 
We assume samples in the consistent group all have the same proportion vector $\bm \alpha = (\alpha_1, \alpha_2, \dots, \alpha_J)^\prime\ \text{with}\ \sum^J_{j=1}\alpha_j=1$, while samples not in the consistent group \ww{can each} have different isoform proportions $\bm \beta^{(d)} = (\beta_1^{(d)}, \beta_2^{(d)}, \dots, \beta_J^{(d)})^\prime$ $\sum^J_{j=1}\beta_j^{(d)}=1$. 
Thus the isoform proportions can be expressed as
\begin{eqnarray*}
\bm \tau^{(d)} & = &  E_d\cdot {\bm \alpha} + (1-E_d) \cdot {\bm \beta}^{(d)}\\
&=&
\left\{
\begin{array}{ll}
\bm \alpha, & \text{if}\ E_d = 1\,, \\
{\bm \beta}^{(d)}, & \text{if}\ E_d = 0\,.
\end{array}
\right.
\end{eqnarray*}
The isoform proportion vector of the consistent group $\alpha$ is our parameter of interest.

\begin{figure}[tb!]
\begin{center}
\includegraphics[trim= 0cm 13cm 0cm 0cm, width=0.7\textwidth]{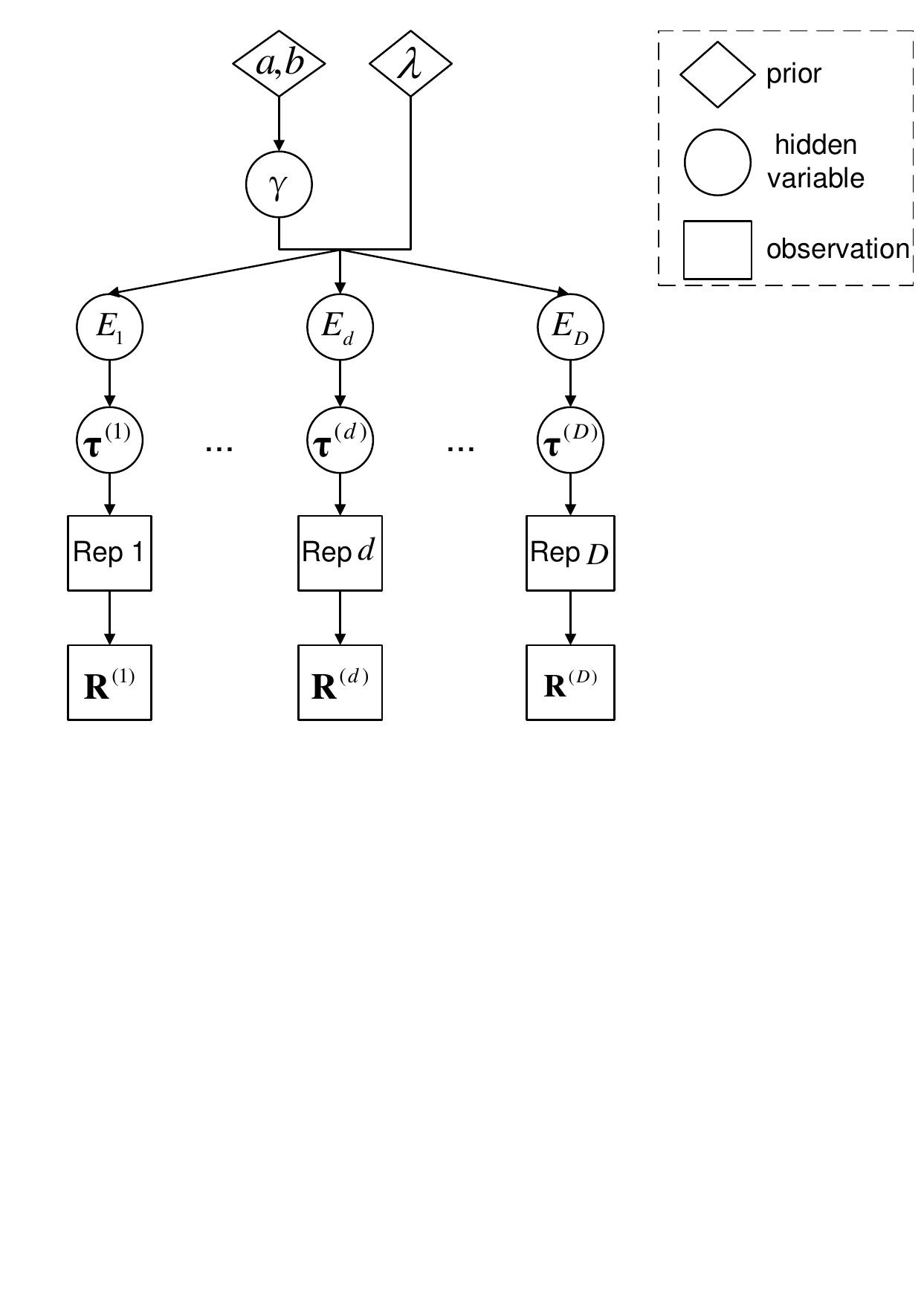}
\end{center}
\caption{
Joint modeling of multiple RNA-seq samples. In this framework, $E_d\ (d=1,2,\dots,D)$ is a binary hidden state variable indicating whether RNA-seq sample $d$ is in the consistent group, while $a,b$ and $\gamma$ are hyper-parameters \JL{(priors)} in $E_d$'s distribution. Depending on $E_d$, the isoform proportion vector $\bm \tau^{(d)}$ takes either the consistent group's isoform proportion vector $\bm \alpha$ or its own $\bm \beta^{(d)}$. Given the isoform proportions, RNA-seq reads are generated in each sample, and our observed data are summarized as \JL{$\bm R^{(d)}$} (see Section \ref{sec:bin}).
\label{fig:model}}
\end{figure}

We assume $\bm \alpha$ and $\bm \beta^{(d)}$ are \textit{a priori} $\text{Dirichlet}(\bm \lambda)$,
and $E_d$ is \textit{a priori} $\text{Bernoulli}(\gamma)$: $E_d |\gamma \sim \text{Bernoulli}(\gamma), \text{where}\ \gamma \sim \text{Beta}(a,b).$
Intuitively,
$\bm \lambda$ controls the distance between the isoform proportions of samples \ww{inside and outside} the consistent group, while
 $\gamma$ controls the tendency of assigning a sample to the consistent group.
We describe the relationship between observed RNA-seq reads and hidden isoform proportions in multiple samples under a Bayesian framework (Fig \ref{fig:model}).

\subsection{The MSIQ model} \label{sec:likfun}
We introduce $I^{(d)}_{i,j}$ as a short notation of \WL{binary variable} $\1\{Z^{(d)}_{i}=j\}$. Then given a sample with isoform proportion $\bm \tau^{(d)}$, the probability of  read $r^{(d)}_{i}$ and origin $Z^{(d)}_{i}$ can be written as \ww{follows}:

\begin{eqnarray}\label{eq:sing_read}
P\left( r^{(d)}_{i} , Z^{(d)}_{i} | \bm \tau^{(d)} \right)
&=&
\prod_{j=1}^J P\left( r^{(d)}_{i}, Z^{(d)}_{i} = j | \bm \tau^{(d)} \right)^{\1\{Z^{(d)}_{i}=j\}} \nonumber \\ 
&=&
\prod_{j=1}^J \left[P\left( r^{(d)}_{i} |  Z^{(d)}_{i} = j\right)\tau^{(d)}_j\right] ^{I^{(d)}_{i,j}}  \triangleq 
\prod_{j=1}^J \left(h_{i,j}^{(d)} \tau^{(d)}_j\right)^{I^{(d)}_{i,j}}, 
\end{eqnarray}
where \JL{$P\left( r^{(d)}_{i} , Z^{(d)}_{i} | \bm \tau^{(d)} \right)$ refers to the joint density of read $r_i^{(d)}$ and its isoform origin $Z_i^{(d)}$ given the model parameters, and} $h_{i,j}^{(d)}$ is the generating probability of read $r_i^{(d)}$ given isoform $j$. If read $r_i^{(d)}$ and isoform $j$ are incompatible (e.g., read 2 in Fig \ref{fig:intro_gene} cannot come from isoform 1), $h_{i,j}^{(d)}=0$. Otherwise, $h_{i,j}^{(d)}$ depends on the model for the read generation mechanism.
We adopt the following model \ww{from} \cite{zhang2014wemiq}\ww{:}
\begin{eqnarray*}
h_{i,j}^{(d)}
= \frac{1}{\ell^\prime_j} \times P\left( L^{(d)}_{i,j}\right),
\end{eqnarray*}
\WL{
where $\ell^\prime_j$ is the effective length (\ww{i.e.,} the number of possible starting positions on the fragment) of isoform $j$ and can be calculated as $\ell^\prime_j = \ell_j- L^{(d)}$: $\ell_j$ is the length of isoform $j$ and $L^{(d)}$ is the mean fragment length in sample $d$.
$ L^{(d)}_{i,j}$ denotes the fragment length of $r^{(d)}_{i}$ if it comes from isoform $j$.
Note that the same read may correspond to different fragment lengths if they come from different isoforms. For example, read 1 in Fig \ref{fig:intro_gene} \ww{corresponds to fragments of different lengths in isoforms 1 and 2.}
$ L^{(d)}_{i,j}$ is assumed to be a Gaussian random variable and its mean $L^{(d)}=\mathbb{E}({L^{(d)}_{i,j}})$ and variance $\text{var}({L^{(d)}_{i,j}})$ can be estimated from single-isoform genes, whose mapped reads directly determine fragment lengths.
}

Let $\bm E = (E_1, E_2, \dots, E_D)^\prime$
be the hidden state vector indicating whether each sample is among the consistent group or not, and let $\bm R=\{\bm R^{(d)}\}_{d=1}^D$, $\bm Z=\{\bm Z^{(d)}\}_{d=1}^D$, and $\bm {\bm \tau}=\{{\bm \tau}^{(d)}\}_{d=1}^D$ represent the reads, origins of reads, and isoform proportions in all the samples\ww{,} respectively. 
To simplify the notation, we also introduce $n^{(d)}_{j}=\sum_{i=1}^{n_d} I^{(d)}_{i,j}$ to represent the total number of reads coming from isoform $j$ in sample $d$.
Given equation (\ref{eq:sing_read}), the joint probability of all reads in \ww{the} MSIQ model is as follows\ww{:}
\begin{eqnarray*}
P\left(\bm R, \bm Z, \bm \tau, \bm E, \gamma | \bm \lambda, a, b\right) = 
P\left( \left. \bm R, \bm Z \right| \bm \tau, \bm E \right)
P\left(\left. \bm \tau \right| \bm\lambda, \bm E \right)
P\left(\left. \bm E \right| \gamma \right)
P(\gamma | a,b)\,,
\end{eqnarray*}
where
\begin{eqnarray*}
P\left( \left. \bm R, \bm Z \right| \bm \tau, \bm E \right)
&=&
\prod_{d=1}^D\left\{
\left[ \prod_{i=1}^{n_d} \prod_{j=1}^J \left( h_{i,j}^{(d)} \alpha_j \right) ^{I^{(d)}_{i,j}}\right]^{E_d}
\left[\prod_{i=1}^{n_d} \prod_{j=1}^J
\left( h_{i,j}^{(d)} \beta^{(d)}_j\right)^{I^{(d)}_{i,j}}
\right]^{1-E_d} 
\right\},
\\
P\left(\left. \bm \tau \right| \bm\lambda , \bm E\right)
&\propto&
\prod_{j=1}^J \alpha_j^{\lambda_j-1} 
\prod_{d=1}^D \left[\prod_{j=1}^J \left( \beta^{(d)}_j  \right)^{\lambda_j-1}\right]^{1-E_d}\,,
\\
P\left(\left. \bm E \right| \gamma \right)
&\propto&
\gamma^{\sum^D_{d=1}E_d} (1-\gamma)^{D-\sum^D_{d=1}E_d}\,,
\\
P(\gamma | a,b)
&\propto&
\gamma^{a-1}(1-\gamma)^{b-1}\,.
\end{eqnarray*}
As a result, the joint probability can be simplified as
\begin{eqnarray}\label{eq:main}
&&P\left(\bm R, \bm Z, \bm \tau, \bm E, \gamma | \bm \lambda, a, b\right) \\
&\propto&
\left[\prod_{j=1}^J
\alpha_{j}^{ \lambda_j -1 +\sum_{d=1}^D E_d n^{(d)}_{j} }\right]
\left[
\prod^D_{d=1}\prod^J_{j=1}
\left(\beta_{j}^{(d)}\right)^{ (1-E_d) \left(\lambda_j -1 +n^{(d)}_{j}\right)}
\right] \nonumber\\
&&
\left[\prod^D_{d=1}\prod^J_{j=1}\prod^{n_d}_{i=1}
\left(h_{i,j}^{(d)} \right) ^{I^{(d)}_{i,j}}\right]
\gamma^{\sum^D_{d=1}E_d + a - 1} (1-\gamma)^{D- \sum^D_{d=1}E_d + b -1}\,.  \nonumber
\end{eqnarray}

\subsection{Markov chain Monte Carlo}
In the MSIQ model (\ref{eq:main}), the reads $\bm R$ are the observed data, the isoform origins $\bm Z$ and the consistent group indicator $\bm E$ are the hidden data, while isoform proportions $\bm \alpha, \{\bm \beta^{(d)}\}^{D}_{d=1}$, and consistent group proportion $\gamma$ are the parameters.
To estimate the \ww{parameters}, a useful approach is to implement a Gibbs sampler to iteratively draw posterior samples of hidden data and parameters from their conditional distributions.
Since our ultimate parameter of interest is $\bm \alpha$, whose inference becomes obvious given $\bm Z$ and $\bm E$,  we integrate out $\bm {\bm \tau}$ (i.e., $\bm \alpha$ and $\{\bm \beta^{(d)}\}^{D}_{d=1}$) in model (\ref{eq:main}) to achieve better computational efficiency. This step is based on \ww{a} property of \ww{the} Dirichlet distribution:
$$
\int \cdots \int_{ \{ (\tau_1,\ldots,\tau_J): 0 \le \tau_j \le 1, \sum_j \tau_j = 1 \} }
\prod_{j=1}^J \tau_j^{\lambda_j -1} d\tau_1\cdots d\tau_J= B(\bm \lambda), \quad \forall \lambda_j > 0\,,
$$
where $B(\bm \lambda)=\frac{\Pi^J_{j=1}\Gamma(\lambda_j)}{\Gamma(\sum^J_{j=1}\lambda_j)}$.
Hence\ww{,}
\begin{eqnarray*}
P\left( \left. \bm R,
\bm Z,
\bm E,  \gamma \right|\bm\lambda, a, b \right)
&\propto&  B_1(\bm Z,
\bm E) \cdot \prod_{d=1}^D
B_0^{(d)}(\bm Z^{(d)},
E_d)
\cdot
\left[
\prod_{d=1}^D
\prod_{i=1}^{n_d}
\prod_{j=1}^J
\left(h_{i,j}^{(d)} \right) ^{I^{(d)}_{i,j}}
\right]\\
&&
\cdot
 \gamma^{\sum^D_{d=1}E_d + a - 1} (1-\gamma)^{D- \sum^D_{d=1}E_d + b -1}\,,
\end{eqnarray*}
where
\begin{eqnarray*}
B_1(\bm Z, \bm E)
&=& \frac{\prod_{j=1}^J \Gamma\left(\lambda_j + \sum_{d=1}^D E_d \cdot n^{(d)}_{j}\right)}{\Gamma\left(\sum_{j=1}^J \lambda_j + \sum_{d=1}^D E_d \cdot n_d  \right)}\,, \\
B_0^{(d)}(\bm Z^{(d)}, E_d=1)
&=& 1,\\
B_0^{(d)}(\bm Z^{(d)}, E_d=0)
&=& \frac{\prod_{j=1}^J \Gamma\left(\lambda_j + n^{(d)}_{j}\right)}{\Gamma\left(\sum_{j=1}^J \lambda_j + n_d \right)}\,.
\end{eqnarray*}

We denote $\Theta=\{\bm Z, \bm E, \gamma\}$. The distribution of each parameter \ww{or} hidden variable conditional on everything else can thus be estimated  by Gibbs sampling as follows.
\begin{itemize}
\item[(1)] $E_{d}$ follows a Bernoulli distribution:
\begin{eqnarray}\label{eq:E_d}
E_{d} | \Theta / \{E_{d}\}
&\sim&
\text{Bern}\left( \frac{odds(E_{d}; \bm \lambda, \tau)}{1+odds(E_{d}; \bm \lambda, \tau)} \right)\,,
\end{eqnarray}
where
\begin{eqnarray*}
odds(E_{d}; \bm \lambda, \tau)
&=&\frac{P(E_{d}=1| \Theta / \{E_{d}\} )}{P(E_{d}=0| \Theta / \{E_{d}\} )}=
\frac
{ P\left( \left. \bm R,
\bm Z,
\bm E_{-d}, E_{d}=1,  \gamma \right|\bm\lambda, a, b \right) }
{ P\left( \left. \bm R,
\bm Z,
\bm E_{-d}, E_{d}=0,  \gamma \right|\bm\lambda, a, b \right) }\\
&=&
\frac{B_1(\bm Z,
\bm E_{-d}, E_{d} = 1)}{B_1(\bm Z,
\bm E_{-d}, E_{d} = 0)}
\cdot
\frac{B_0^{(d)}(\bm Z^{(d)}, E_d=1) }{B_0^{(d)}(\bm Z^{(d)}, E_d=0) }
\cdot
\frac{\gamma}{1-\gamma}\,.
\end{eqnarray*}
\item[(2)] $Z^{(d)}_i$ follows a multinomial distribution:
\begin{eqnarray}\label{eq:Z}
Z^{(d)}_i | \Theta / \{Z^{(d)}_i\}
&\sim&
\text{Multinomial}\left(q^{(d)}_{i,1}, q^{(d)}_{i,2}, \dots, q^{(d)}_{i,J} \right)\,,
\end{eqnarray}
where
 \begin{eqnarray*}
q^{(d)}_{i,j} &=&
\frac{P(Z^{(d)}_i=j | \Theta / \{Z^{(d)}_i\} )}
{\sum^J_{j^\prime=1}P(Z^{(d)}_i=j^\prime | \Theta / \{Z^{(d)}_i\} )}
=\frac{P\left(\left.  \bm R,
\bm Z^{(-d)}_{-i} , Z^{(d)}_i = j,
\bm E,  \gamma \right|\bm\lambda, a, b \right)}
{\sum^J_{j^\prime=1}P\left(\left.  \bm R,
\bm Z^{(-d)}_{-i} , Z^{(d)}_i = j^\prime,
\bm E,  \gamma \right|\bm\lambda, a, b \right)}
\,.
\end{eqnarray*}

\item[(3)] $\gamma$ follows a Beta distribution:
\begin{eqnarray}\label{eq:gamma}
\gamma | \Theta / \{\gamma\}
\sim \text{Beta}\left(\sum_{d=1}^D E_d + a, D-\sum_{d=1}^D E_d + b\right)\,.
\end{eqnarray}
\end{itemize}

\subsection{Estimators of isoform \ww{proportions}}\label{sec:estimator}
With the above posterior distribution of the hidden variables and parameters, we can draw samples iteratively to estimate the hidden state of each RNA-seq sample and the true isoform \ww{proportions} in the consistent group. Suppose we have $T$ iterations available after discarding the burn-in period of Gibbs sampling.
In each iteration, we denote the sampled hidden state vector as $\bm E^{(t)}=(E^{(1)}_{1}, E^{(2)}_{2}, \dots, E^{(T)}_{D})^\prime$ and the hidden origin vector in sample 
$d$ as $(Z^{(d,t)}_1,\dots,Z^{(d,t)}_{n_d})^\prime$.

To estimate isoform \ww{proportions} in each iteration, we pool the reads from sample $d$ whose state varibale $E^{(t)}_{d} = 1$ to calculate $\bm \alpha^{(t)}$, where
\begin{eqnarray}\label{eq: alpha_j}
\bm \alpha_j^{(t)} = 
\frac{\lambda_j+\sum^D_{d=1}\left(E_d^{(t)} \sum^{n_d}_{i=1}\1\{Z^{(d,t)}_i=j\}\right)}{\sum^J_{j=1}\lambda_j+\sum^D_{d=1}E_d^{(t)}n_d}.
\end{eqnarray}
Overall, the MSIQ estimator of the isoform  \ww{proportions} becomes
\begin{eqnarray*}
\hat{\bm \alpha}^\text{MSIQ} = \frac{1}{T}\sum_{t=1}^T\bm \alpha^{(t)},
\end{eqnarray*}
and the relative estimation error is calculated as
$$\text{REE}(\hat{\bm \alpha}^\text{MSIQ}) = \sum^m_{j=1}|\alpha_j - \hat{\bm \alpha}_j^\text{MSIQ}|/\alpha_j.$$
\WL{We \JL{prove} the consistency property of the MSIQ estimator $\hat{\bm \alpha}^{MSIQ}$ in the following lemma. (Please refer to the supplementary information for the complete proof.)
\begin{lemma}\label{ourlemma}
$\hat{\bm \alpha}^\text{MSIQ}$ converges to the posterior mean of isoform proportion $\mathbb{E}(\bm \alpha|\bm R, \bm \lambda, a, b)$:
$$\lim_{T\to \infty}\hat{\bm \alpha}^\text{MSIQ}=\mathbb{E}(\bm \alpha|\bm R, \lambda, a, b) .$$
\end{lemma}
}

We can also estimate the posterior probability of each sample belonging to the consistent group: $\bm \theta=(\theta_{1}, \theta_{2}, \dots, \theta_{D})^\prime$, where $\theta_{d} = P(E_d=1|\bm R, \bm \lambda, a, b)$, and the estimator is
$$\hat\theta_{d}^\text{MSIQ} = \frac{1}{T}\sum_{t=1}^T E^{(t)}_{d}.$$
\ww{Based on this posterior probability, we predict the state variable of each sample: $\hat{E}_{d}=\1\{\hat\theta_{d}^\text{MSIQ} > 1/2\}$}.

\ww{To further evaluate the biological variation within the consistent group, we estimate the standard error of the MSIQ estimator given the posterior samples drawn by the Gibbs sampling. For isoform $j$, the standard error of the respective entry in $\hat{\bm \alpha}^\text{MSIQ}$ is estimated as:
\begin{align}\label{eq:std}
\hat\sigma_j = \sqrt{\frac{1}{T}\sum^T_{t=1}(\bm \alpha_j^{(t)} - \hat{\bm \alpha}_j^\text{MSIQ})^2}\,.
\end{align}
Note that the consistent group is automatically selected by the MSIQ model given the overall heterogeneity among samples. Even though the consistent group is assumed to have a consensus isoform proportion, it is useful to account for the biological variation, especially when the overall heterogeneity is non-negligible.
}

We also consider six competing estimators  to demonstrate the effectiveness of MSIQ in accurate isoform quantification.
From what has been derived in Section \ref{sec:likfun}, we know that the log likelihood of all reads in sample $d$ is
\begin{eqnarray*}
\log\left(P( \bm R^{(d)}, \bm Z^{(d)} | \bm \tau^{(d)} ) \right)
&=&
\sum_{i=1}^{n_d}\sum_{j=1}^m I^{(d)}_{i,j}
\log\left(h^{(d)}_{i,j} \tau^{(d)}_j\right).
\end{eqnarray*}
Then the EM algorithm can be implemented to estimate $\bm \tau^{(d)}$ by maximizing the log likelihood. The six competing estimators are calculated using the EM algorithm based on different sets of samples:
\begin{description}
\item{\textbf{AVG}} (averaging): We calculate the isoform proportion in each sample and take the average of them as the estimator of isoform proportion\ww{,} 
$$\hat{\bm \alpha}^{\text{AVG}}=\frac{1}{D}\sum^D_{d=1}\hat{\bm \tau}^{(d)}.$$
\item{\textbf{AVG*} (oracle averaging)}: We calculate the isoform proportion in each sample in the true consistent group and take the average of them as the estimator of isoform proportion\ww{,}
$$\hat{\bm \alpha}^{\text{AVG*}}=\frac{\sum^D_{d=1}\hat{\bm \tau}^{(d)}\1\{E_d=1\}}{\sum^D_{d=1}\1\{E_d=1\}}.$$
\item{\textbf{POOL}} (pooling): We pool the reads in all samples together, then we use the EM algorithm to estimate the isoform proportion $\bm \tau$ as $\hat{\bm \alpha}^{\text{POOL}}$.
\item{\textbf{POOL*}} (oracle pooling): We pool the reads in samples in the true consistent group together, then we use the EM algorithm to estimate  $\bm \tau$ as $\hat{\bm \alpha}^{\text{POOL*}}$.
\item{\textbf{MSIQa}} (MSIQ averaging): We calculate the isoform proportion in each sample in the \WL{consistent group (identified by MSIQ)} and take the average of them as the estimator of isoform proportion\ww{,}
$$\hat{\bm \alpha}^{\text{MSIQa}}=\frac{\sum^D_{d=1}\hat{\bm \tau}^{(d)}\1\{\hat\theta_{d}^\text{MSIQ} > 1/2\}}{\sum^D_{d=1}\1\{\hat\theta_{d}^\text{MSIQ} > 1/2\}}.$$
\item{\textbf{MSIQp}} (MSIQ pooling): We pool the reads of the given gene in the samples in the \WL{consistent group (identified by MSIQ)}  together, then we use the EM algorithm to estimate  $\bm \tau$ as $\hat{\bm \alpha}^{\text{MSIQp}}$.
\end{description}

Among these estimators, $\hat{\bm \alpha}^{\text{AVG*}}$ and $\hat{\bm \alpha}^{\text{POOL*}}$ are oracle estimators that we take as gold standards in simulations but are unknown in real data; $\hat{\bm \alpha}^{\text{MSIQa}}$ and $\hat{\bm \alpha}^{\text{MSIQp}}$ are MSIQ-dependent and rely on $\hat{\bm\theta}$ estimated by MSIQ. 


\section{Results}\label{sec:results}

\subsection{Performance of MSIQ in simulations}\label{sec:simu}
To show that MSIQ provides more accurate estimates \ww{of} isoform expression than the current averaging or pooling method, we compare the relative estimation errors (REE) of $\hat{\bm \alpha}^\text{MSIQ}$ with those of the six competing estimators: $\hat{\bm \alpha}^\text{AVG*}$, $\hat{\bm \alpha}^\text{MSIQa}$, $\hat{\bm \alpha}^\text{AVG}$, $\hat{\bm \alpha}^\text{POOL*}$, $\hat{\bm \alpha}^\text{MSIQp}$,  and $\hat{\bm \alpha}^\text{POOL}$. 
It is difficult to compare these methods on real data because  true isoform abundances in samples are unknown.
Although the quantitative polymerase chain \ww{reaction} (qPCR) technology can accurately measure the abundance of mRNA isoforms and produce ``gold standard'' isoform abundance, qPCR data sets are scarce and unavailable for most biological conditions \citep{li2011rsem}.
We use simulated data to compare the \ww{performances} of these estimators under various scenarios and parameter settings.

\begin{table}[!bt]
\caption{Four parameter settings and five scenarios in the simulation study.
\label{tab:simulation}}
\begin{center}
\begin{tabular}{ccc}
\hline
setting & average fragment length (bp) & read length (bp) \\
\hline
1 & 150 & 50  \\
2 & 250 & 50  \\
3 & 150 & 100  \\
4 & 250 & 100 \\
\hline
scenario & \% samples in the consistent group & isoform proportions\\
\hline
1 & 100& $\{\bm \alpha, \bm \alpha, \bm \alpha, \bm \alpha, \bm \alpha,
          \bm \alpha, \bm \alpha, \bm \alpha, \bm \alpha, \bm \alpha\}$\\
2 & 50& $\{\bm \alpha, \bm \alpha, \bm \alpha, \bm \alpha, \bm \alpha,
          \bm \beta_1, \bm \beta_2, \bm \beta_3, \bm \beta_4,  \bm \beta_5\}$\\
3 & 70& $\{\bm \alpha, \bm \alpha, \bm \alpha, \bm \alpha, \bm \alpha,
          \bm \alpha, \bm \alpha, \bm \beta_1, \bm \beta_2, \bm \beta_3\}$\\
4 & 70& $\{\bm \alpha, \bm \alpha, \bm \alpha, \bm \alpha, \bm \alpha,
          \bm \alpha, \bm \alpha, \bm\beta_6, \bm\beta_6, \bm\beta_6\}$\\
5 & 70& $\{\bm \alpha, \bm \alpha, \bm \alpha, \bm \alpha, \bm \alpha,
          \bm \alpha, \bm \alpha, \bm\beta_7, \bm\beta_7, \bm\beta_7\}$\\
\hline
\end{tabular}
\end{center}
\end{table}

We simulate RNA-seq reads from $3,421$ \emph{D.melanogaster} (fly) genes that have \WL{multiple isoforms} in the annotation (September 2010) \ww{available in the} UCSC Genome Browser.
\ww{Among these genes, $221$ have $3$ exons, $330$ have $4$ exons, $365$ have $5$ exons, $370$ have $6$ exons, $320$ have $7$ exons, $311$ have $8$ exons, $256$ have $9$ exons, $292$ have $10$ exons, and $956$ genes have more than 10 exons.}
The isoform numbers increase at a roughly exponential rate as the exon numbers increase
 (see Supplementary Fig S2A). We simulate \ww{ten} samples and $500$ paired-end reads from each gene in every sample. 
To fully evaluate the \ww{performances} of the seven estimators, we consider five different scenarios with different numbers of samples in the consistent group.

For each gene, we first independently generate the isoform proportion vector $\bm \alpha$ for the samples in the consistent group and the isoform proportion vectors $\bm \beta_1, \bm \beta_2, \bm \beta_3, \bm \beta_4\, \text{and}\ \bm \beta_5$ for \ww{the} other five samples. The five scenarios are designed as follows (see Table \ref{tab:simulation}).
\begin{itemize}
\item In scenario 1, all \ww{ten} samples are in the consistent group.
\item In scenario 2, \ww{five} samples are in the consistent group, and the other \ww{five} samples have individual isoform proportions $\bm \beta_1, \bm \beta_2, \bm \beta_3, \bm \beta_4\ \text{and}\ \bm \beta_5$.
\item In scenario 3, \ww{seven} samples are in the consistent group, and the other \ww{three} samples have individual isoform proportions $\bm \beta_1, \bm \beta_2\ \text{and}\  \bm \beta_3$.
\item In scenario 4, \ww{seven} samples are in the consistent group, and the other \ww{three} samples have the same isoform proportion vector as $$\bm\beta_6=\argmax\limits_{\bm\beta_i, i=1,\ldots, 5} || \bm\beta_i - \bm\alpha ||_2^2\,,$$ which is the isoform proportion vector most different from $\bm \alpha$.
\item In scenario 5, \ww{seven} samples are in the consistent group, and the other \ww{three} samples have the same isoform proportion vector as $$\bm \beta_7=\argmin\limits_{\bm\beta_i, i=1,\ldots, 5} || \bm\beta_i - \bm\alpha ||_2^2\,,$$ which is the isoform proportion vector \ww{most} similar to $\bm \alpha$.
\end{itemize}

We also consider four settings of fragment and read length (Table \ref{tab:simulation}) to examine how these parameters affect the performance\ww{s} of the seven estimators on isoform quantification. Under each setting, we first determine the origin of a fragment according to the designated isoform proportion, and then the starting position and the fragment length can be simulated from a uniform distribution and a normal distribution\ww{, respectively} (with a standard deviation of $10$ bp). Once the \ww{starting} and ending positions of the fragments are determined, the corresponding paired-end reads are also obtained.

For each scenario and parameter setting, we calculate the seven estimators, and then evaluate their estimation accuracy by calculating the REE of these estimates against the true isoform proportions. 
\WL{When calculating $\hat{\bm \alpha}^\text{MSIQ}$, we set the hyper-parameters in model (\ref{eq:main}) as $a=7$ and $b=2$. We have also included a sensitivity analysis of the MSIQ method on these two parameters in the supplementary information.}

\begin{figure}[!hbt]
\begin{center}
\includegraphics[width=\textwidth]{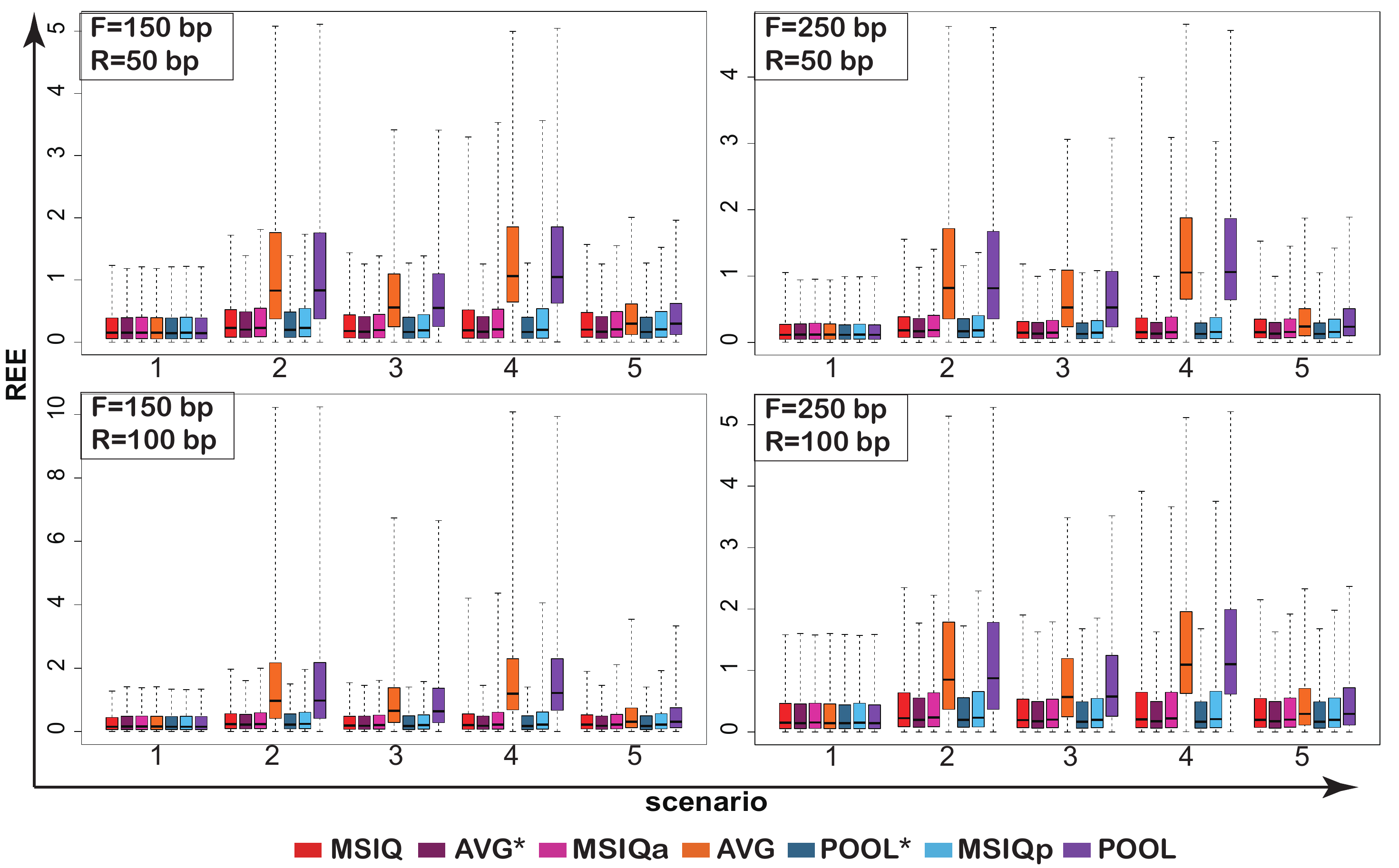}
\end{center}
\caption{
Relative estimation error (REE) rates of the seven estimators in \ww{scenarios} 1-5. REE rates are calculated on 2,465 fly genes with 3-10 exons. In each boxplot, the REE rates of \ww{MSIQ, AVG$^*$ (oracle averaging), MSIQa, AVG (averaging), POOL$^*$ (oracle pooling), MSIQp and POOL (pooling)} are plotted side by side under each scenario and the whiskers extend to the most extreme REE rates. The top-right legend of each plot displays the parameter setting: the mean fragment length (F) and the read length (R).
\label{fig:simu_box}}
\end{figure}

\subsubsection{MSIQ achieves the lowest error rates in different scenarios}
We calculate the error rates of the seven estimators for the $2,465$ fly genes with no more than \ww{ten} exons in different scenarios and parameter settings, and illustrate the results in Fig \ref{fig:simu_box}.
 The results suggest that given the samples not in the consistent group (scenarios 2-5), especially when these samples constitute a large proportion or are vastly different from the consistent group, MSIQ ($\hat{\bm \alpha}^\text{MSIQ}$) and MSIQ-based methods ($\hat{\bm \alpha}^\text{MSIQa}$\ \text{and}\ $\hat{\bm \alpha}^\text{MSIQp}$) achieve much smaller error rates than the averaging or pooling \ww{methods} ($\hat{\bm \alpha}^\text{AVG}$ and $\hat{\bm \alpha}^\text{POOL}$). 
Compared with $\hat{\bm \alpha}^\text{MSIQ}$, $\hat{\bm \alpha}^\text{AVG}$ results in a $17.3$-fold increase in the REE rates on average\ww{,} and $\hat{\bm \alpha}^\text{POOL}$ results in a $17.6$-fold increase.
We also summarize the REE of the seven estimators (see Fig \ref{fig:simu_box2} in the Appendix) when we include the $956$ fly genes with more than \ww{ten} exons.
The isoform quantification task is much more challenging for these $956$ genes since they have \ww{many} more annotated isoforms (see Supplementary Fig S2A). 
As expected, both the largest and the average REE rates increase with the addition of these 956 genes, because their complicated isoform structures \ww{introduce} more difficulty and complexity \ww{in} model fitting and computation.
These results suggest that\ww{,} compared with the direct averaging or pooling method, the MSIQ methods\ww{, which} take the quality of samples into consideration\ww{,} can lead to more accurate isoform quantification when multiple RNA-seq samples are available. 
Fig \ref{fig:simu_box} also shows that MSIQ can constrain the estimation error to a much narrower range compared with direct averaging and pooling. 
MSIQ is able to control the REE rate below $1.33$ for 90\% of the $2,465$ genes, while direct averaging and pooling give rise to REE rates larger than $2.00$ for more than 15\%  of these genes. We conclude that MSIQ is a more robust method than direct averaging and pooling.

\begin{table}[!bt]
\caption{Median REE rates of five estimators under the five scenarios. The values are averaged over the four parameter settings and rounded to three decimal places. Differences in REE rates between MSIQ and the four other estimators are listed in parentheses.
\label{tab:med_err}}
\begin{center}
\begin{tabular}{lccccc}
\hline
estimator & scenario 1 & scenario 2 & scenario 3 & scenario 4 & scenario 5 \\
\hline
MSIQ & 0.157& 0.236& 0.194& 0.208& 0.211 \\
AVG* & 0.158& 0.215& 0.179& 0.179& 0.179 \\
         &  (-0.001)&  (0.021)& (0.014)&  (0.029)&  (0.031) \\
MSIQa & 0.164& 0.244& 0.202& 0.222& 0.217 \\
      & (-0.007)& (-0.009)& (-0.009)& (-0.014)& (-0.006) \\
POOL* & 0.152& 0.212& 0.175& 0.175& 0.175 \\
         & (0.005)&  (0.023)&  (0.019)&  (0.033)&  (0.036) \\
MSIQp & 0.157& 0.242& 0.200& 0.217& 0.215 \\
       & (-0.000)& (-0.006)& (-0.007)& (-0.009)& (-0.005) \\
\hline
\end{tabular}
\end{center}
\end{table}

\begin{figure}[bt!]
\begin{center}
\includegraphics[width=\textwidth]{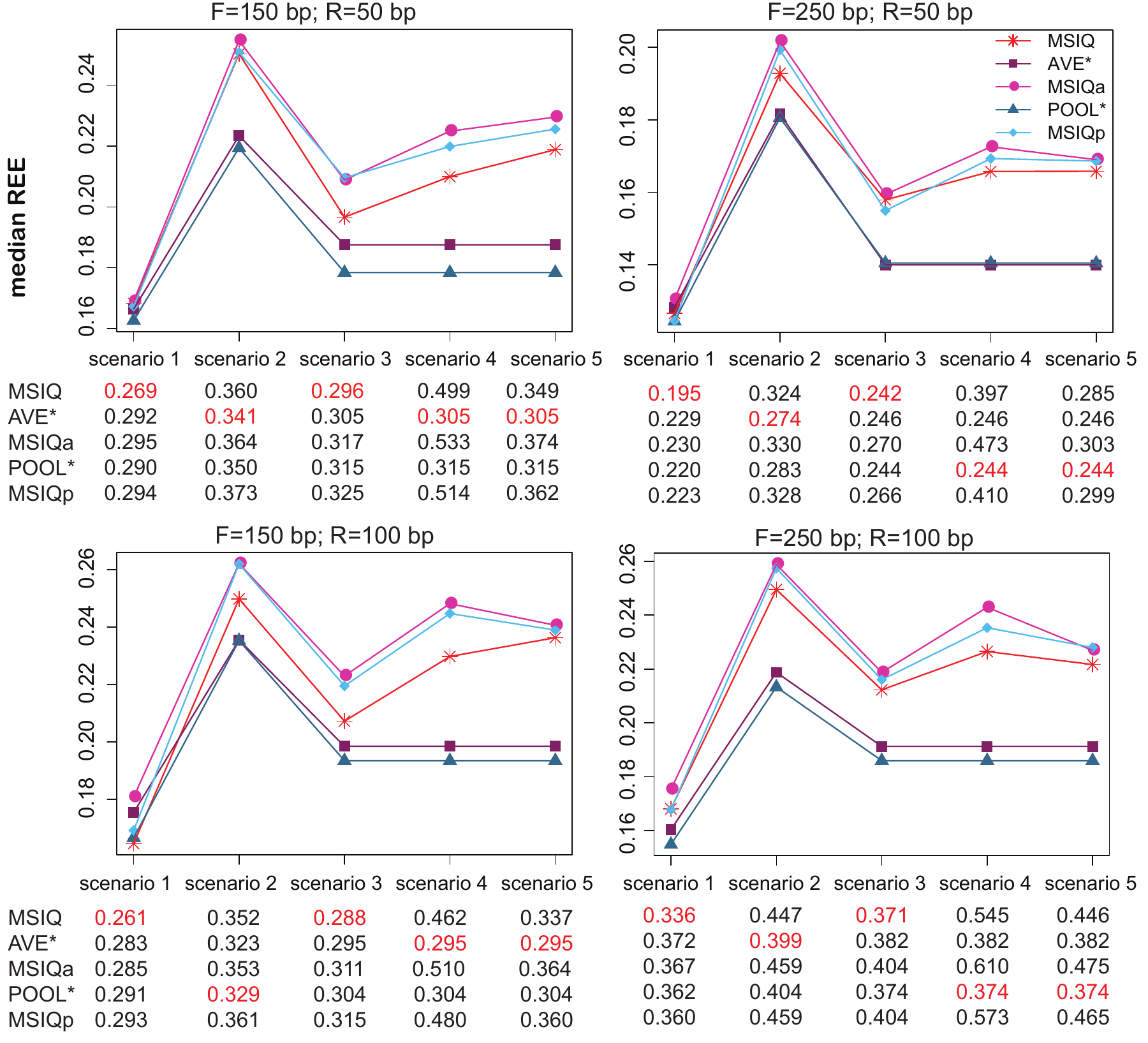}
\end{center}
\caption{
Median REE rates of the MSIQ-based and oracle estimators in scenarios 1-5. MSIQ outperforms MSIQa and MSIQp and gives error rates close to those of the oracle estimators. \ww{The parameter setting: the mean fragment length (F) and the read length (R) are listed on the top of each panel. The standard errors of MSIQ's REE rates are given under each scenario. The smallest standard error in each scenario is marked in red.}
\label{fig:simu_med}}
\end{figure}

We also summarize the median REE of these estimators under different \ww{scenarios} in Fig \ref{fig:simu_med} and Table \ref{tab:med_err}. The results show that MSIQ not only outperforms direct averaging and pooling\ww{,} as we have seen, but also achieves more accurate abundance estimation than MSIQa and MSIQp.  Compared with MSIQ's median REE rate, MSIQa and MSIQp have \ww{average REE rates that are greater by 0.009 and 0.007,} respectively.
From Fig \ref{fig:simu_med} and Table \ref{tab:med_err} we also conclude that the estimation results of MSIQ are similar to those of AVG$^*$ and POOL$^*$, the two oracle estimators that are impossible to calculate on real data. On average, the REE rate of MSIQ is only 0.019 larger than AVG$^*$ and 0.058 larger than POOL$^*$.

\subsubsection{Different scenarios influence estimators' performance}
Since AVG and POOL are observed to have much poorer accuracy than the other five estimation methods, we remove them from the comparison for a more detailed evaluation of the other five methods.
From Fig \ref{fig:simu_med}, it is obvious that the proportion of samples in the consistent group and the difference between the consistent group and other samples have large effects on the performances of all five estimating methods: MSIQ, AVG*, MISQa, POOL*, and MISQp.
In scenario 1 when all the samples are in the consistent group, the five methods \ww{exhibit their} lowest median REE rates \ww{for the} $2,465$ genes.
In scenario 2, which has the smallest proportion of samples in the consistent group, all five methods have the largest median REE rates among all scenarios. This phenomenon can be explained by the fact that \ww{having} fewer samples in the consistent group \ww{leads} to more error-prone identification of these samples and less accurate estimates of the isoform proportions.
In scenarios 3, 4, and 5, \ww{in which} 70\% \ww{of the samples are} in the consistent group, the REE rates of \ww{the} five methods lie between those of scenarios 1 and 2. 
Among all three non-oracle estimation methods (MSIQ, MSIQa and MSIQp), MSIQ has the best performance in all five scenarios.
Unlike MSIQa and MSIQp, which discard the samples outside of the identified consistent group, MSIQ partially borrows information from these samples through the Bayesian hierarchical framework.

\begin{figure}[bt!]
\begin{center}
\includegraphics[width=\textwidth]{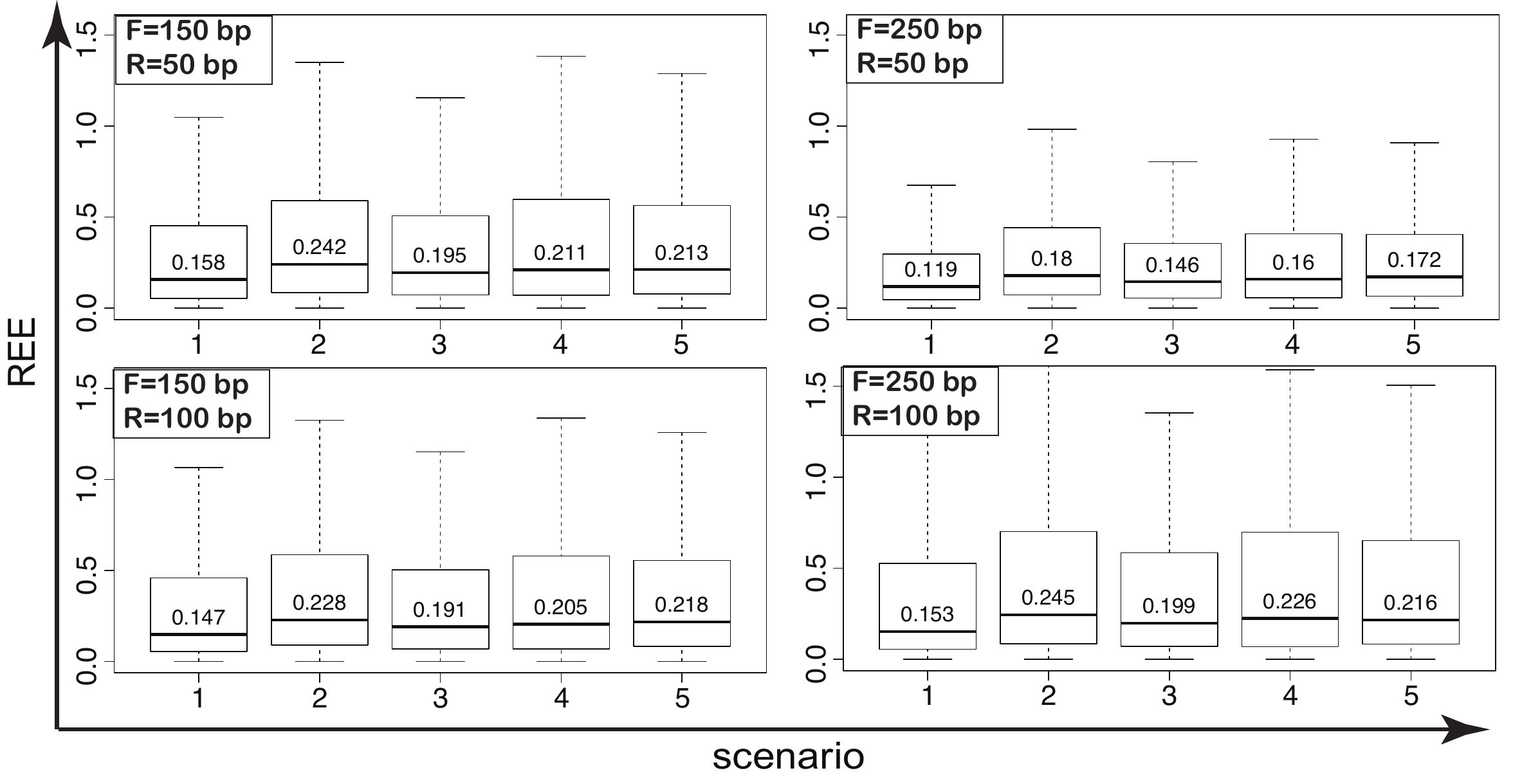}
\end{center}
\caption{REE rates of MSIQ for RNA-seq samples with different fragments and read lengths. \ww{The median, the 1st quartile and the 3rd quartile of the REE rates in different scenarios are illustrated in the boxplots}, respectively. The \ww{top left} legend of each plot displays the parameter setting: the mean fragment length (F) and the read length (R).\label{fig:simu_length}}
\end{figure}

\subsubsection{More accurate isoform quantification with longer fragments}
We also evaluate the REE rates of MSIQ with different fragment lengths and read lengths in simulated RNA-seq experiments.
The 1st quartile, median, and 3rd quartile of the REE errors in each of the five scenarios are illustrated in Fig \ref{fig:simu_length}.
It is obvious that longer fragment lengths would improve the estimation accuracy, especially when read lengths are short.
Specifically, when read lengths are set to 50 bp, increasing fragment lengths from 150 to 250 bp leads to a 22.5\% decrease in the median REE rate and a 31.8\% decrease in the inter-quartile range of REE; when read lengths are set to 100 bp, the increase of fragment lengths does not make as much difference.

\subsection{Performance of MSIQ on real data}\label{sec:realdata}
\subsubsection{MSIQ has \ww{the} highest estimation accuracy in a pseudo real data study}\label{sec:real_hESC}
Although the true isoform proportions are mostly unknown in real data, 
we are still able to evaluate multi-sample isoform abundance estimation methods
by creating a set of samples with the majority from one tissue of interest (the consistent group) and other samples from a different tissue.
Even though this setup is not a realistic scenario in biological studies, it provides a good opportunity to evaluate different \ww{estimation} methods. In this setup, we know the true states of the hidden state \ww{variables}, i.e., which samples belong to the consistent group. If our MSIQ method \ww{performs well}, its estimated isoform proportions on all the samples should be close to its estimates on the samples in the consistent group only.
We use six public RNA-seq data sets of human embryonic stem cells (hESC) and consider these samples \ww{to be} the consistent group. We mix these samples with three samples of human brain tissues or three samples simulated by Flux Simulator \citep{griebel2012modelling}. Please see Supplementary Table S2 for detailed description. 

\begin{table}[!bt]
\caption{Description of RNA-seq samples \ww{inside and outside} the consistent group in five sets.
\label{tab:sample}}
\begin{center}
\begin{tabular}{cccccc}
\hline
set ID & consistent group & other samples & sample IDs \\
\hline
1  & hESC & / & 1-6\\
2  & hESC & brain & 1-9\\
3  & hESC & Flux Simulator & 1-6, 10-14\\ 
4  & hESC & Flux Simulator & 1-6, 15-19\\
5  & hESC & Flux Simulator & 1-6, 20-24\\
\hline
\end{tabular}
\end{center}
\end{table}

We obtain five sets of RNA-seq samples by mixing the six hESC samples in the consistent group with other samples in different combinations (Table \ref{tab:sample}). 
Because MSIQ has the best performance among all the three non-oracle MSIQ-based estimation methods (\ww{i.e.,} MSIQ, MSIQa and MSIQp) in the simulation studies in Section \ref{sec:simu}, we only consider MSIQ \ww{and not MSIQa or MSIQp} in the real data studies.
We compare MSIQ with direct averaging (AVG) and pooling (POOL) on these five sets of real RNA-seq samples to estimate the isoform proportions in the consistent group (hESC).
We also consider three previously developed methods for single RNA-seq samples \ww{(i.e., Cufflinks, MISO and iReckon)} in this comparison. For Cufflinks, we use both the averaging (Cuffa) and the pooling (Cuffp) approach to calculate the isoform \ww{proportions}. For MISO and iReckon, pooling is not a feasible approach due to \ww{the} extremely large memory \ww{requirements when analyzing} a merged RNA-seq sample with a huge size, so we only consider the averaging approach.
When evaluating the above seven methods, we consider each method's estimates  on set 1 as the standards, because set 1 only contains the six hESC samples (\ww{i.e.,} the consistent group).
The estimation results of MSIQ, AVG, POOL, Cuffa, Cuffp, MISO and iReckon on sets 2 \ww{through} 5 are compared with their own standard on set 1, and REE rates are calculated accordingly.

In our study, the true mRNA isoform structures are extracted from the \emph{Homo sapiens} annotation (February 2009) of the UCSC Genome Browser \citep{rosenbloom2015ucsc}. 
According to the annotation, there are $15,268$ human genes with multiple isoforms.
Supplementary Fig S2B summarizes the distribution of the numbers of exons and isoforms of these genes.
We can see that the isoform structures of \ww{humans} are much more complex \ww{than those of} simple model \ww{organisms} like \ww{fruit flies}.
For each sample set, we only perform estimation for genes that have reads in all the samples. As a result, isoform proportions are calculated for $11,091$ genes in set 1, $9753$ genes in set 2,  $460$ genes in set 3, $404$ genes in set 4, and $497$ genes in set 5.

\begin{figure}[!bt]
\begin{center}
\includegraphics[width=\textwidth]{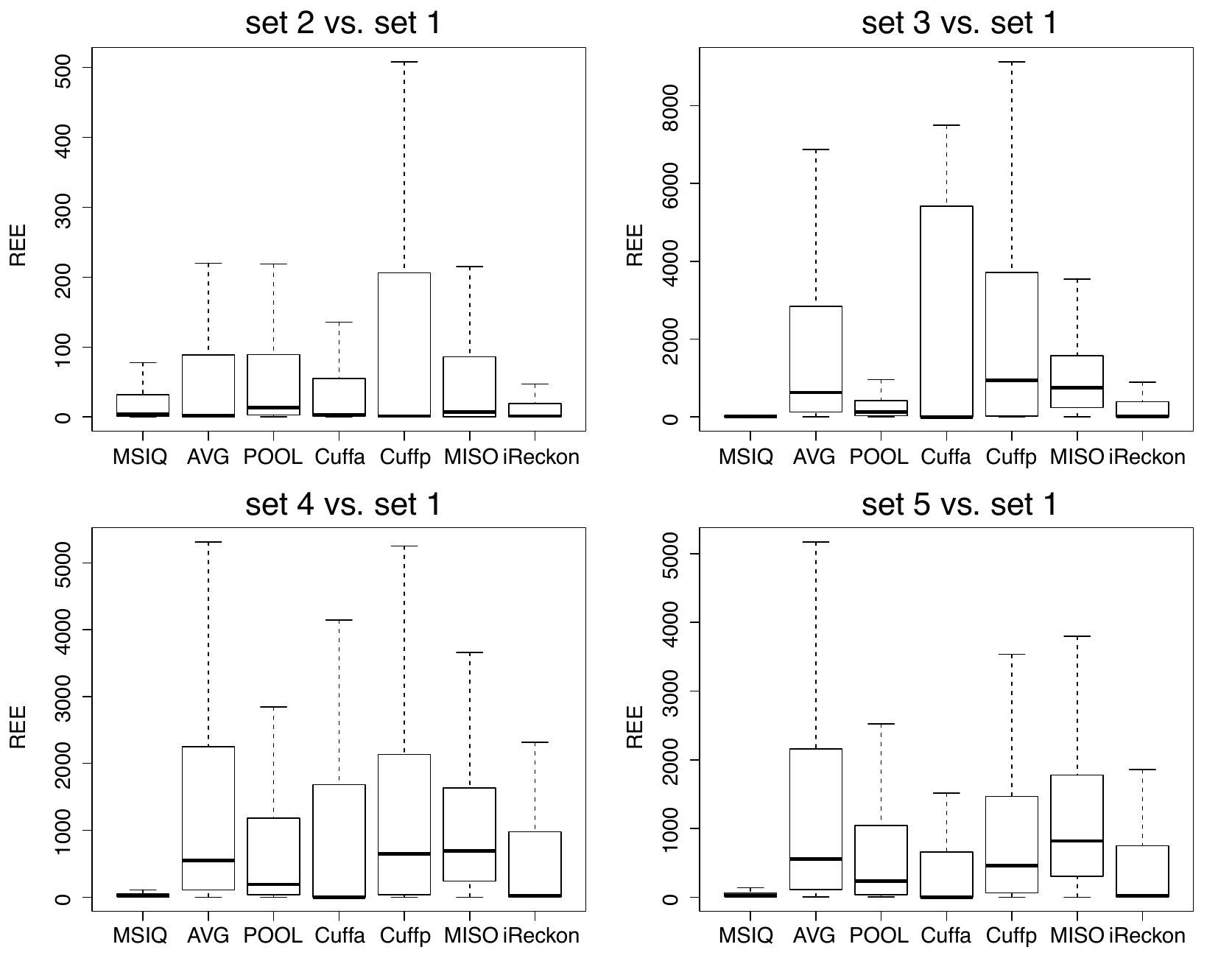}
\end{center}
\caption{
\WL{REE rates of MSIQ, AVG (averaging), POOL (pooling), Cuffa (Cufflinks averaging), Cuffp (Cufflinks pooling), MISO and iReckon on sets 2 to 5. We use these seven estimators to perform isoform quantification on sets 2 to 5 and calculate REE by treating their correpsonding estimates on set 1 as the standards.}
\label{fig:real_box}}
\end{figure}

Comparing the REE rates of MSIQ and the other six methods in Fig \ref{fig:real_box}, we clearly see that MSIQ generally achieves the lowest median error rates and the smallest inter-quantile ranges in all the four comparison cases. This result is strong evidence supporting the effectiveness of MSIQ in identifying the consistent group and estimating its isoform proportions.
Note that even though iReckon also leads to relatively accurate results, especially in set 2 vs. set 1, the number of genes for which iReckon can provide estimation is much smaller compared with other methods. In the four cases, iReckon obtains \ww{estimates} only for $1065$, $255$, $377$ and $374$ genes. 
This comparison also suggests that pooling is not an ideal approach when the depths of sequencing coverage in multiple RNA-seq samples vary \ww{greatly}.

\begin{figure}[!bth]
\begin{center}
\includegraphics[width=\textwidth]{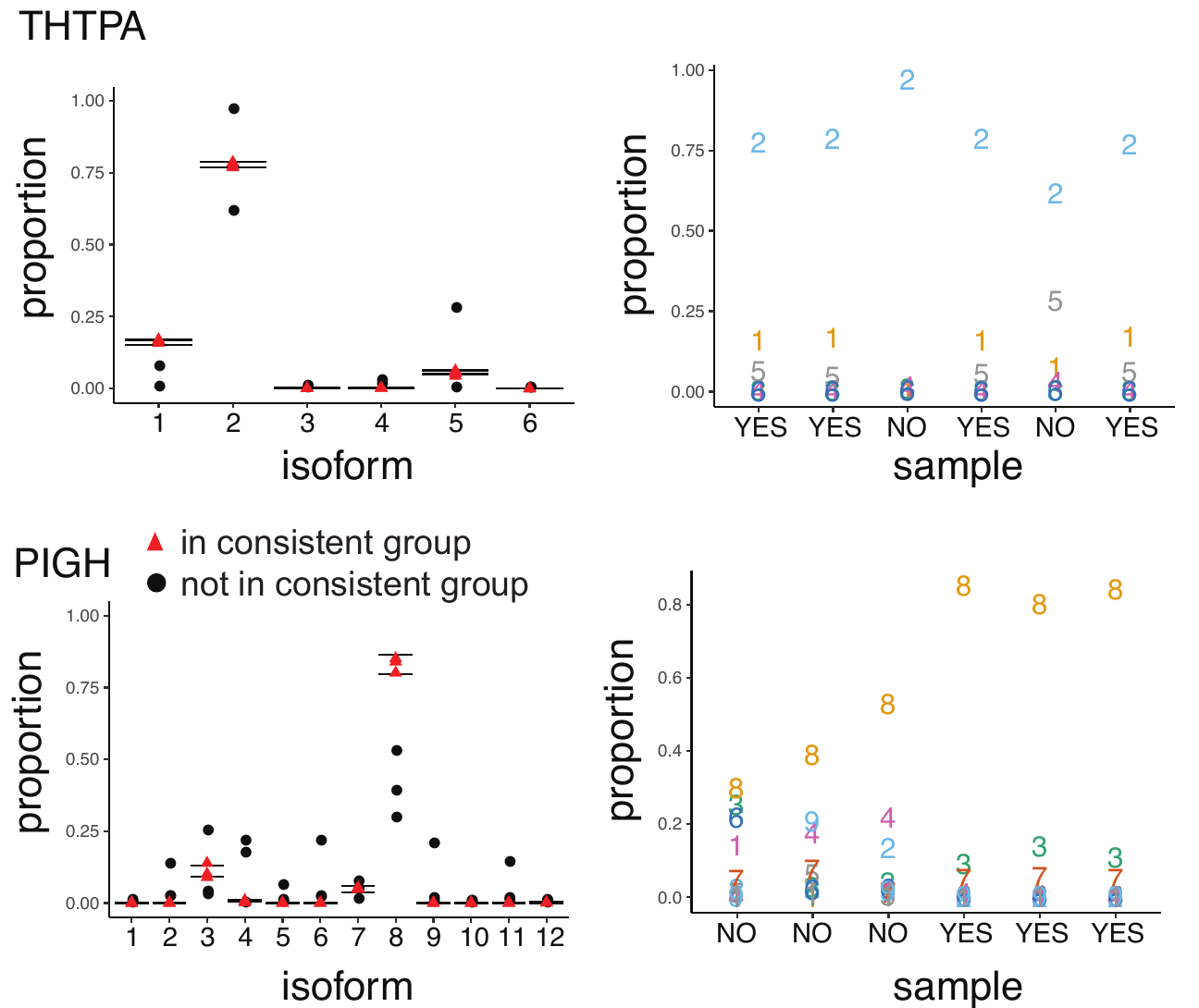}
\end{center}
\caption{\ww{MSIQ's estimated isoform proportions and standard errors for gene \textit{THTPA} ($6$ isoforms) and gene \textit{PIGH} ($12$ isoforms). The left plots give the estimated isoform proportions by isoform. The intervals are the respective MSIQ estimator $\pm$ one standard error: $\hat\alpha_j^{\text{MSIQ}}\pm \hat\sigma_j$. The right plots give the estimated isoform proportions by sample. The numbers denote the isoform indices and the horizontal axis denotes whether the corresponding sample is identified as being within the consistent group or not.
}
\label{fig:variation}}
\end{figure}

\ww{We also use set 1 (i.e., the $6$ hESC samples) in this study to illustrate why the consistent group represents more reliable transcriptome landscapes and how the standard deviation defined in formula (\ref{eq:std}) can be used to assess the biological variation within the consistent group. Shown in Figure \ref{fig:variation} are two example genes \textit{THTPA} ($6$ isoforms) and \textit{PIGH} ($12$ isoforms). We use these two examples to illustrate that (1) MSIQ is bale to identify \ww{consistent groups} that have comparably more consistent isoform abundances\ww{, and} (2) the biological variation within the consistent group is much smaller compared to the overall variation among all the samples, and this variation is well captured by the estimated standard errors.
}

\subsubsection{\WL{MSIQ leads to \ww{the highest} correlation with NanoString counts}}\label{sec:real_liver}
We present a second real data example to evaluate different methods by comparing their reported isoform abundances (in FPKM values) with \ww{NanoString} counts on the same data sets. The \ww{NanoString} nCounter technology is considered \ww{to be} a highly reproducible and robust method for detecting gene and isoform expression \citep{kulkarni2011digital}. As a consequence, the \ww{NanoString} measurements are widely used as a benchmark for isoform expression \citep{germain2016rnaonthebench,steijger2013assessment}. We compare our \ww{MSIQ method} with three other estimation methods\ww{,} Cufflinks, iReckon, and MISO\ww{,} based on their performances on six samples of \ww{the} human HepG2 (liver hepatocellular carcinoma) immortalized cell line (see Supplementary Table S3 for detailed description).

\begin{figure}[!bt]
\begin{center}
\includegraphics[width=\textwidth]{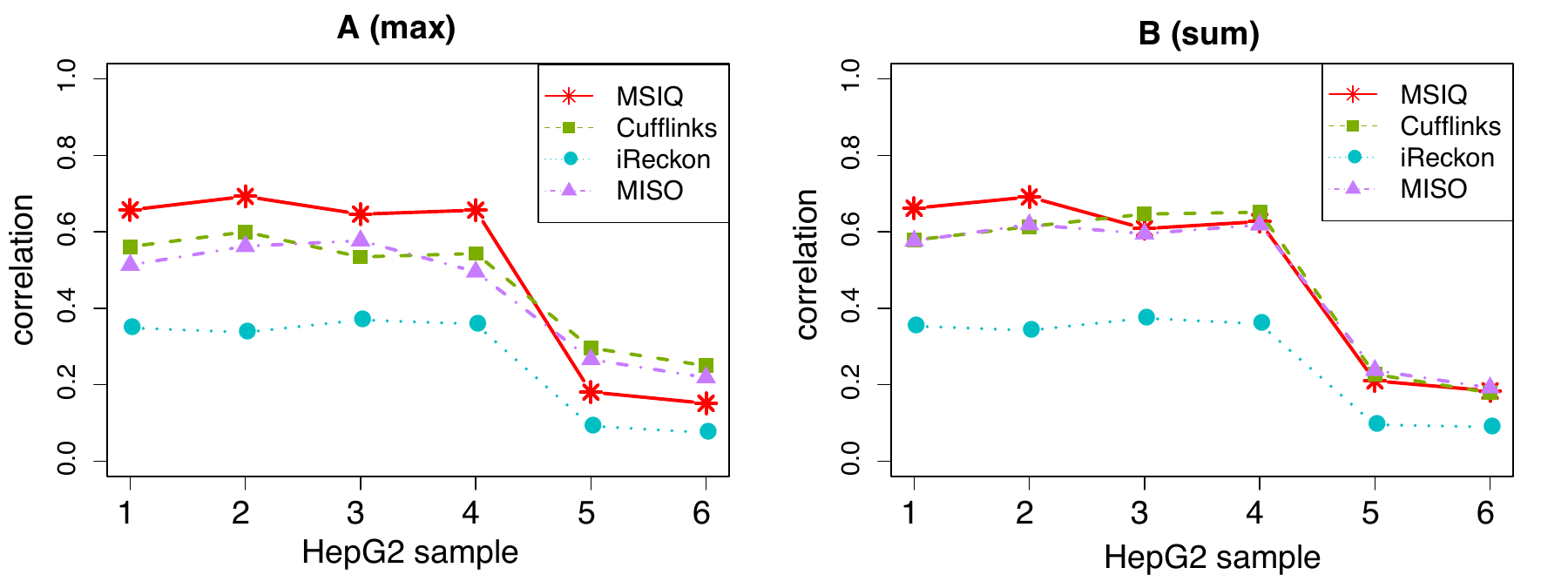}
\end{center}
\caption{
Correlation between \ww{NanoString} counts and the estimated isoform expression. \textbf{A}: For each \ww{NanoString} probe, the corresponding isoform with \ww{the} largest estimated FPKM value is used to calculate the correlation. \ww{The standard error of the calculated correlation coefficients is between $0.069$ and $0.099$.} \textbf{B}: For each \ww{NanoString} probe, the sum of all the corresponding isoforms' estimated FPKM values is used to calculate the correlation. \ww{The standard error of the calculated correlation coefficients is between $0.065$ and $0.085$.}
\label{fig:liver_corr}}
\end{figure}

Even though genome-wide isoform abundances are not available for these HepG2 data, the \ww{NanoString} counts are available for a small set of genes \citep{steijger2013assessment}. These \ww{NanoString} measurements include $140$ probes that correspond to $470$ isoforms in $107$ genes.  We apply MSIQ, Cufflinks, iReckon and MISO on the six HepG2 samples and use each method to estimate isoform abundances for this set of genes. Cufflinks and iReckon directly report the FPKM values of the relevant isoforms. MSIQ and MISO estimate isoform proportions, and the FPKM values can be calculated accordingly. 
For each sample, we calculate the Pearson correlation coefficient between each method's estimated isoform expression and the benchmark \ww{NanoString} counts. Since the \ww{NanoString} probe counts do not have \ww{a} one-to-one correspondence with isoform expression, for each \ww{NanoString} probe we either use the isoform with \ww{the} largest expression (Fig \ref{fig:liver_corr}A) or add up the expression of all the isoforms (Fig \ref{fig:liver_corr}B).
Overall, the estimated expression of MSIQ has \ww{the highest} correlation with the \ww{NanoString} counts and achieves \ww{the} best consistency with this benchmark measurement, compared with Cufflinks, iReckon and MISO. Please note that samples 5 and 6 are found not belonging to the consistent group by MSIQ, and that \ww{is} why MSIQ does not have the highest correlations on them. This observation is coherent with the definition of \ww{a} consistent group by MSIQ. This result again suggests that MSIQ leads to more accurate isoform quantification by incorporating the information in multiple RNA-seq samples.

\section{Discussion and conclusion}
\label{sec:conc}
In this paper, we propose a new method\ww{,} MSIQ\ww{,} to more accurately estimate isoform expression levels \ww{associated with biological conditions of interest using} multiple RNA-seq data \ww{sets}.
Accurate isoform quantification from RNA-seq data has long been a challenge because the existence of multiple isoforms makes it impossible to uniquely assign many reads and determine the reads' isoform origins.
MSIQ tackles this challenge by utilizing data from multiple RNA-seq samples \ww{derived from} the same biological condition\ww{; we reason} that aggregating more information can improve accuracy in isoform abundance estimation.
Unlike previous work that treats all the samples equally, MSIQ identifies \ww{a} consistent group of samples that are most representative of the biological condition and estimates isoform proportions of the consistent group.

Applications of MSIQ to both simulated and real data demonstrate that MSIQ \ww{yields} more accurate isoform quantification than direct averaging or pooling methods given the existence of \ww{poor} quality or mislabeled samples. These results suggest MSIQ's potential as a powerful and robust transcriptomic tool for isoform expression quantification. 
\ww{MSIQ's estimation results provide robust and accurate transcriptome profiles, which can be used to construct co-expression networks, investigate cell-type-specific isoform expression, and identify differentially expressed transcripts between two biological conditions.}
\ww{The MSIQ method also provides standard error estimates to measure the variability of isoform abundance within the consistent group. This information can be especially useful when users need to compare  multiple tissue or cell types.
We estimate the standard errors using the posterior samples of isoform proportions, and we note that our method can be extended to directly model the variability parameters at the cost of increased complexity in the model and computations.
In addition to isoform abundance estimation, MSIQ can also be applied to evaluate the quality of multiple RNA-seq samples of the same tissue or cell type. This application can help researchers evaluate the reproducibility of RNA-seq samples and determine which samples to include in downstream analyses.}

An important step in our MSIQ method is the identification of the consistent group, which depends on posterior draws of the hidden state variables. 
We currently use a Beta-Bernoulli model to describe the probability of each sample belonging to the consistent group. \ww{However}, it is possible to improve the model once gold standard data (i.e., qPCR) for  the \ww{biological condition of interest} become available \citep{adamski2014method,li2011rsem}.
We can extend our MSIQ model to account for the heterogeneous quality of multiple RNA-seq samples based on the similarity of the isoform abundance estimates \ww{between} each sample and the gold standard.
Such quality assessment can be integrated with the inter-sample similarity to better identify the consistent group.
As a result, the samples that have higher agreement with  gold  standards and high similarity with each other will be more likely to be considered \ww{a part of} the consistent group.  This procedure is supposed to identify more reliable samples and \ww{can} potentially increase the re-use of public RNA-seq data as it will provide an interpretable measure of the quality of multiple RNA-seq data sets. We would also like to point out that \ww{biological knowledge can be incorporated into MSIQ modeling} to further improve isoform abundance estimation. For example, mRNA fragments are\ww{, in fact,} not uniformly distributed within the isoforms \citep{zhang2014wemiq}, and \ww{a} high correlation was observed between read coverage and genome GC content \citep{li2011sparse}. Our proposed hierarchical model can be considered an umbrella framework that can be easily extended to incorporate more detailed modeling procedures as long as these procedures use likelihoods to describe read generating processes. Such extension might help MSIQ achieve better performance on complex genes.

Another interesting extension of our MSIQ method is to model single-cell RNA-seq (scRNA-seq) data, which contain information on the technical and biological noise of isoform abundance at the single-cell level \citep{wu2014quantitative,macaulay2014single}.
scRNA-seq data are \ww{needed for} the analysis of (1) subpopulations of cells from a larger heterogeneous population and (2) rare cell types, for which sufficient material cannot be obtained for conventional RNA-seq experiments \citep{mortazavi2008mapping}.
Given scRNA-seq data \ww{on} multiple cells from the same population, MSIQ can be iteratively utilized to evaluate the transcriptional heterogeneity and detect subpopulations (\ww{i.e.,} consistent groups) in the set of samples. Meanwhile, MSIQ can also reveal the principal isoform expression pattern in \ww{a} given cell population. An alternative approach is to allow \ww{for multiple consistent groups as} subpopulations of single cells in statistical modeling. 

The RNA-seq data sets used in the paper are all publicly available. Their accession numbers are provided in the supplementary information. The MSIQ method is implemented in the R package \verb|MSIQ|, which is freely available at \url{https://github.com/Vivianstats/MSIQ}.

\section*{Acknowledgments}
Dr. Jingyi Jessica Li was supported by the start-up fund of the UCLA Department of Statistics and the Hellman Fellowship. Dr. Shihua Zhang was supported by the National Natural Science Foundation of China, No. 61379092, 61422309, the Outstanding Young Scientist Program of CAS and the Key Laboratory of Random Complex Structures and Data Science, CAS (No. 2008DP173182). We thank Dr. Yucheng Yang for his help in the data processing and Dr. Katherine R. McLaughlin for commenting on our work. The authors would also like to thank the reviewers for their contributions to improve the paper.

\clearpage
\appendix

\section{Figure Appendix}\label{app:fig}
\setcounter{table}{0}
\renewcommand{\thetable}{D\arabic{table}}
\setcounter{figure}{0}
\renewcommand{\thefigure}{A\arabic{figure}}

\begin{figure}[!htb]
\begin{center}
\includegraphics[width=\textwidth]{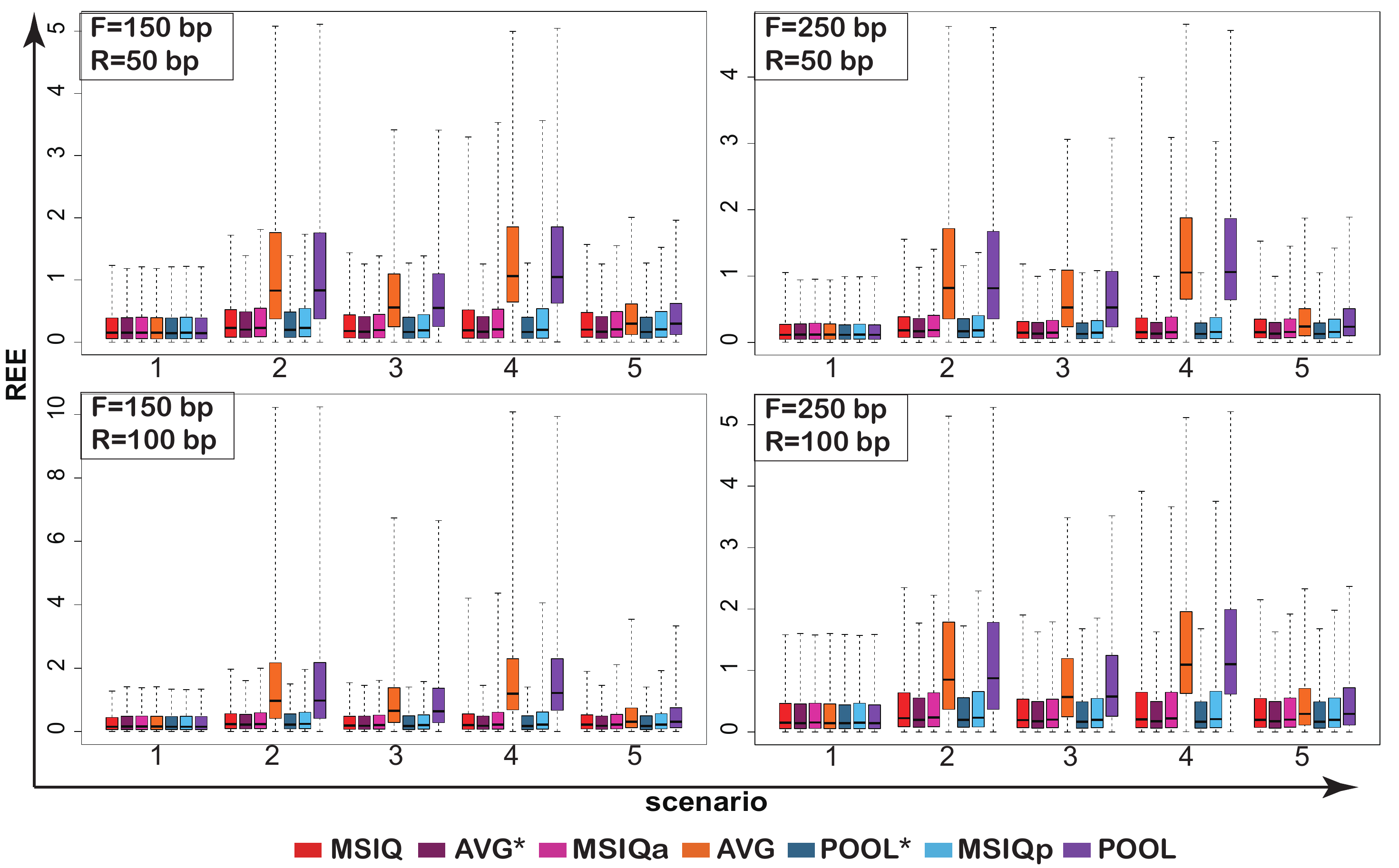}
\end{center}
\caption{
Relative estimation error (REE) rates of the seven estimators in scenario 1-5. REE
rate are calculated on 3,421 fly genes with 3-98 exons. In each boxplot, the REE rates of
MSIQ, AVG*, MSIQa, AVG, POOL*, MSIQp, and POOL are plotted side by side under each
scenario \Rtwo{(with the order of methods listed under the scenario 5 of the bottom left panel)} and the whiskers extend to the most extreme REE rates. The top-right legend
of each plot displays the parameter setting: the mean fragment length (F) and the read
length (R).
\label{fig:simu_box2}}
\end{figure}


\bibliographystyle{Chicago}
\bibliography{MSIQ}

\includepdf[pages=-,pagecommand={},width=1.5\textwidth]{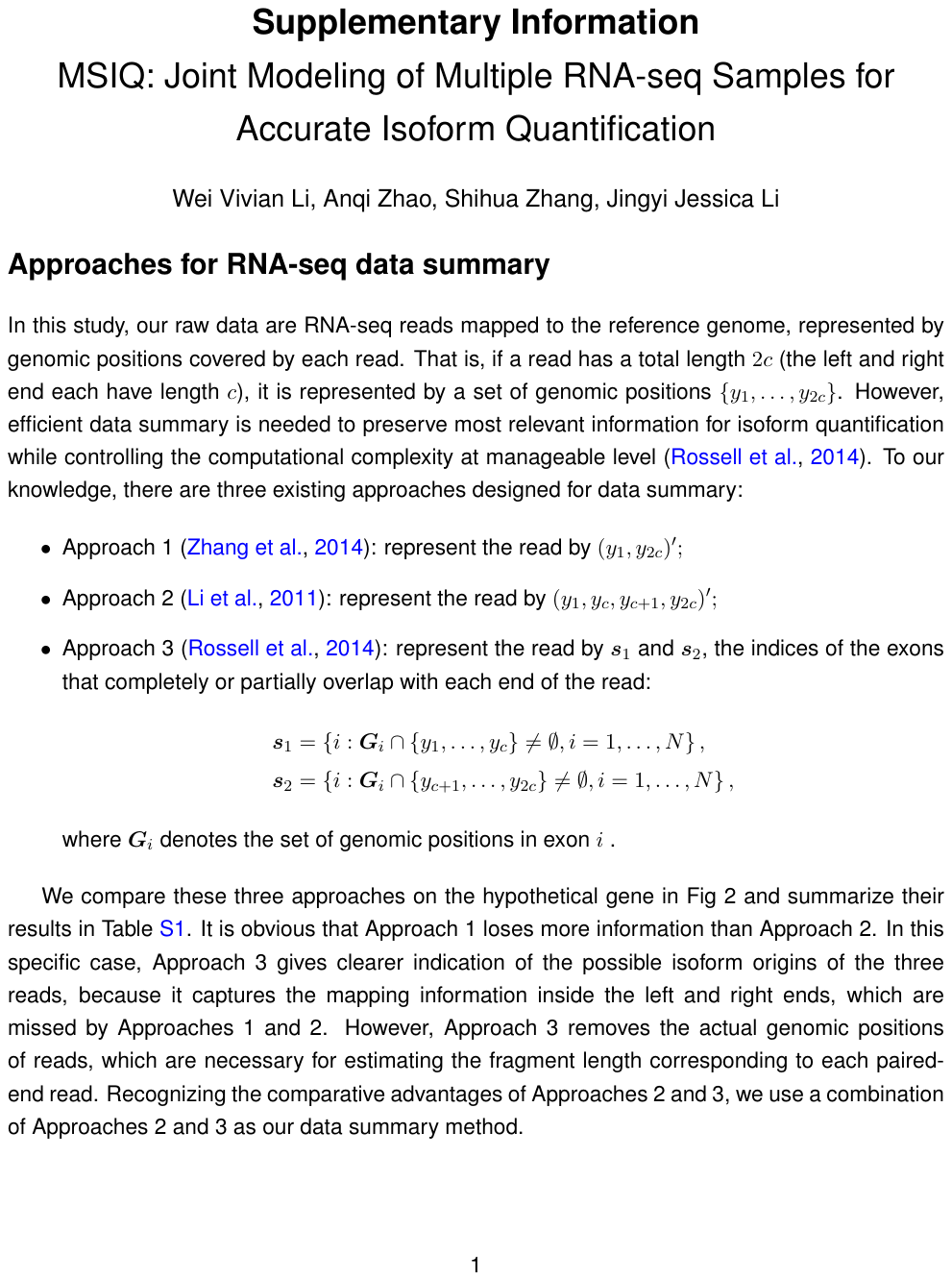}
\end{document}